\documentclass{mn2e}

\usepackage[small]{caption}
\usepackage{graphicx}
\usepackage{epstopdf}
\usepackage{multirow}
\usepackage{amssymb}
\usepackage{float}
\title[A broad-band view of MR\,2251-178]{A broad-band X-ray view of the Warm Absorber in radio-quiet quasar MR\,2251-178.}
\author[J. Gofford et al.]{J. Gofford$^{1}$, J. N. Reeves$^{1,2}$, T. J. Turner$^{2}$, F. Tombesi$^{3,4}$, V. Braito$^{5}$, D. Porquet$^{6}$, \newauthor L. Miller$^{7}$, S. B. Kraemer$^{8}$, Y. Fukazawa$^{9}$ \\
$^{1}$Astrophysics Group, School of Physical Sciences, Keele University, Keele, Staffordshire, ST5 8EH, UK\\
$^{2}$Department of Physics, University of Maryland Baltimore County, Baltimore, MD 21250, USA\\
$^{3}$Department of Astronomy and CREEST, University of Maryland, College Park, MD 20742, USA\\
$^{4}$X-ray Astrophysics Laboratory, NASA/GSFC, Greenbelt, MD 20771, USA\\
$^{5}$Department of Physics and Astronomy, University of Leicester, University Road, Leicester, LE1 7RH, UK\\
$^{6}$Observatoire Astronomique de Strasbourg, Universit\'e de Strasbourg, CNRS, UMR 7550, 11 rue de l'Universit\'e, 67000 Strasbourg, France\\
$^{7}$Department of Physics, University of Oxford, Denys Wilkinson Building, Keble Road, Oxford, OX1 3RH, UK\\
$^{8}$Institute of Astrophysics \& Computational Sciences, Department of Physics, The Catholic University of America, Washington, \\DC 20064, USA\\
$^{9}$Department of Physical Sciences, Hiroshima University, 1-3-1 Kagamiyama, Higashi-Hiroshima, \\Hiroshima 739-8526, Japan}

\begin{document}

\date{Accepted for publication in MNRAS. Pre-edited version.}

\pagerange{\pageref{-}--\pageref{-}} \pubyear{2011}

\maketitle

\label{firstpage}

\begin{abstract}
We present the analysis of a new broad-band X-ray spectrum ($0.6-180.0$\,keV) of the radio-quiet quasar MR\,2251-178 which uses data obtained with both \textit{Suzaku} and the \textit{Swift}/BAT. In accordance with previous observations, we find that the general continuum can be well described by a power-law with $\Gamma=1.6$ and an apparent soft-excess below 1\,keV. Warm absorption is clearly present and absorption lines due to the Fe\,UTA, Fe\,L (Fe\,\textsc{xxiii-xxiv}), S\,\textsc{xv} and S\,\textsc{xvi} are detected below $3$\,keV. At higher energies, Fe\,K absorption from Fe\,\textsc{xxv-xxvi} is detected and a relatively weak ($EW=25^{+12}_{-8}$\,eV) narrow Fe\,K$\alpha$ emission line is observed at $E=6.44\pm0.04$\,keV. The Fe\,K$\alpha$ emission is well modelled by the presence of a mildly ionised ($\xi\la30$) reflection component with a low reflection fraction ($R<0.2$). At least 5 ionised absorption components with 
$10^{20} \la N_{\rm H} \la 10^{23}$\,cm$^{-2}$ and $0 \la \log\xi/{\rm erg\,cm\,s^{-1}} \la 4$ are required to achieve an adequate spectral fit. Alternatively, we show that the continuum can also be fit if a $\Gamma\sim2.0$ power-law is absorbed by a column of $N_{\rm H}\sim10^{23}$\,cm$^{-2}$ which covers $\sim30$\% of the source flux. Independent of which continuum model is adopted, the Fe\,L and Fe\,{\sc xxv} He$\alpha$ lines are described by a single absorber outflowing with $v_{\rm out}\sim0.14$\,c. Such an outflow/disk wind is likely to be substantially clumped ($b\sim10^{-3}$) in order to not vastly exceed the likely accretion rate of the source.
\end{abstract}


\section{Introduction}
\label{intro}
It is now well established that the soft X-ray spectrum of at least half of all Seyfert 1 galaxies is characterised by regions of photoionised ``warm'' absorption along the line of sight (Blustin et al. 2005; McKernan et al. 2007). Extensive studies of the local Seyfert population with \textit{Chandra} and \textit{XMM-Newton} have revealed that the warm absorber (WA) is an intricate array of narrow absorption lines, often seen to be moderately blueshifted with $v\sim100-1000$\,km\,s$^{-1}$, from various ionisation stages of abundant elements such as C, N, O, Ne, Mg, Si, S and Fe (e.g. Kaastra et al. 2000; Kaspi et al. 2002; Crenshaw, Kraemer \& George 2003). Detailed modelling of the WA with photoionisation codes such as \textsc{xstar} (Bautista \& Kallman 2001) have shown that the absorbing material is typified by an ionisation parameter in the range $\log\xi/\rm{erg\,cm\,s}^{-1}\sim 0-3$, a column density of between $N_{\rm H}\sim10^{20}-10^{23}$\,cm$^{-2}$, and most likely originate in a wind outflowing from the putative torus (Blustin et al. 2005) or the later stages of an accretion disk wind (Proga \& Kallman 2004). Energetically, due to their low $v_{\rm out}$, soft X-ray WAs typically have kinetic luminosities to the order of $\sim1\%$ of an AGN's bolometric luminosity (e.g. Blustin et al. 2005) and are therefore unlikely play an important role in terms of AGN feedback scenarios. 

In addition, there is now also a large body of observational evidence for blueshifted absorption lines at rest-frame energies greater than 7\,keV in many AGN (e.g. PG\,1211+143, Pounds et al. 2003; PDS\,456, Reeves et al. 2003; NGC\,1365, Risaliti et al. 2005; MCG\,-5-23-16, Braito et al. 2007; H\,1413+117, Chartas et al. 2007; Mrk\,766, Miller et al. 2007; NGC\,3516, Turner et al. 2008; Mrk\,509, Cappi et al. 2009; 3C 445, Reeves et al. 2010 \& Braito et al. 2011; Tombesi et al. 2010a; Tombesi et al. 2010b). This absorption is generally attributed to Fe\,\textsc{xxv/xxvi} in an outflowing accretion disk wind and requires both a high column density (i.e. $N_{\rm H}=10^{23}-10^{24}$\,cm$^{-2}$ or above), high ionisation parameter (i.e. $\log\xi/\rm{erg\,cm\,s}^{-1}\sim3-6$), and an outflow velocity $\sim0.1$\,c.

Given their large outflow velocities the resulting kinetic power of accretion disk winds can reach a significant fraction of a sources bolometric luminosity (e.g. Pounds \& Reeves 2007; Tombesi et al. 2010b) and they may represent a driving mechanism for AGN feedback processes.  Indeed, extrapolating the kinetic luminosity over a conservative active phase and outflow duty cycle for a given AGN, it is often found that the total mechanical output can be in excess of, or is at least very similar to, the binding energy of a reasonably sized galaxy bulge. Accretion disk winds could therefore be a possible explanation for the observed $M-\sigma$ relation (Ferrarese \& Merritt 2000; Gebhardt et al. 2000); where mass ejected from a radiatively driven accretion  disk wind (e.g. King 2003) can move out into a host galaxy, sweeping up material until, once a critical black hole mass has been reached, the ISM has been evacuated of material and both star formation and SMBH growth cease (see King 2010, and references therein)

\subsection{The Radio-Quiet Quasar MR\,2251-178}
\label{MR2251}
MR\,2251-178 ($z=0.064$; Bergeron et al. 1983; Canizares et al. 1978) was first detected as a bright X-ray source during the \textit{Ariel V} all-sky survey (Cooke et al. 1978) and subsequently identified as a radio-quiet quasar by Ricker et al. (1978) using \textit{SAS-3} observations. The quasar is located on the outskirts of a cluster of approximately 50 galaxies (Phillips 1980) and is surrounded by a large extended nebula of diffuse gas which is characterised by [O\,\textsc{iii}] emission in the optical (Bergeron et al. 1983). More recent observations have also noted similar extended emission in the X-ray band (Kaspi et al. 2004; Gibson et al. 2005). The source is observed to be a  weak radio emitter, and has a Fanaroff-Riley type I (FR\,I) morphology.

The first detailed spectral study of MR\,2251-178 in the X-rays was conducted by Halpern (1984) who noticed that absorption in the soft X-ray band varied on time scales of around $1$ year. The variability implied an order of magnitude increase in the absorption column and was attributed to the change in ionisation of material along the line of sight. Historically, this is regarded as the first suggestion of an X-ray AGN warm absorber. 

Subsequent observations with {\it EXOSAT}, {\it Ginga} and {\it BeppoSAX} established that the broadband X-ray spectrum of MR\,2251-178 can be well described by a power-law of photon-index $\Gamma\sim1.7$ which is absorbed by a column density of around a few\,$\times10^{22}$\,cm$^{-2}$ (Pan et al. 1990; Mineo \& Stewart 1993), and a high-energy roll-over at around $100$\,keV (Orr et al. 2001). Mineo \& Stewart (1993) also found that the ionisation state of the absorbing material was strongly correlated with the source luminosity. 

In the ultra-violet (UV), Monier et al. (2001) found absorption lines due to Ly$\alpha$, N\,\textsc{v} and C\,\textsc{iv} blueshifted in the rest-frame by $\sim300$\,km\,s$^{-1}$. The C\,\textsc{iv} absorption was later shown by Ganguly et al. (2001) to vary over a period of roughly 4 years which enabled the authors to deduce a maximum distance of $\sim$2.4\,kpc between the absorption clouds and the continuum source.

Kaspi et al. (2004), using a series of \textit{ASCA}, \textit{BeppoSAX} and \textit{XMM-Newton} observations which spanned 8.5 years, confirmed that the continuum can be described by an absorbed power-law of photon index $\Gamma\sim1.6$ but found that it further required an additional soft excess at low X-ray energies. The {\it XMM-Newton} spectrum required at least two or three separate ionised absorbers with column densities in the range $10^{20-22}$\,cm$^{-2}$ and had physical properties which appeared to vary between observations. This lead the authors to posit a scenario where absorption clouds were moving across the line of sight over the timescale of `several months'. Further UV absorption lines from C\,\textsc{iii}, H\,\textsc{i} and O\,\textsc{vi} were detected in the \textit{FUSE} spectrum, which were blueshifted with velocities similar to those found by Monier et al. (2001).

MR\,2251-178 has also been observed by the {\it Chandra}/HETG which revealed evidence of a highly-ionised, high-velocity (i.e.\ $v=-12700\pm2400$\,km\,s$^{-1}$) outflow in the Fe\,K band (Gibson et al. 2005). Gibson et al. (2005) attributed this absorption to the Ly$\alpha$ line of Fe\,\textsc{xxvi}, and inferred that unless the absorber has a low covering fraction the mass-loss rate of MR\,2251-178 is at least an order of magnitude larger than the accretion rate.  

In this paper we present a new broad-band X-ray observation of MR\,2251-178, using data obtained with both \textit{Suzaku} and the \textit{Swift} Burst Alert Telescope (BAT). We begin by discussing the data reduction process in \S \ref{data-reduction}, and then parameterise the broadband spectrum in \S \ref{preliminary-fitting}. In \S\ref{absorption-lines} we discuss the several absorption lines which are detected and determine their statistical significance, before performing a detailed modelling of the spectrum with the {\sc xstar} photoionisation code in \S\ref{xstar-modelling}. To calculate luminosities, a concordance cosmology with H$_0=71$ km\,s$^{-1}$\,Mpc$^{-1}$, $\Omega_{\Lambda}$=0.73, and $\Omega_m$=0.27 (Spergel et al. 2003) was adopted.

\begin{table*}
	\begin{center}
\begin{minipage}{140mm}
	\caption{Summary of observation parameters}
		\begin{tabular}{l l c c c c l}
			\hline
			 & \multirow{2}{*}{Instrument} & \multirow{2}{*}{Date} & Exposure & Count rate & Flux & \multirow{2}{*}{Obs. ID} \\
			 &			  &		 & (ks) 	& (s$^{-1}$) & ($\times10^{-11}$\,erg\,cm$^{-2}$\,s$^{-1}$) \\
			\hline\hline
		\multirow{4}{*}{Suzaku} & XIS-FI & \multirow{4}{*}{07/05/09} & \multirow{2}{*}{136924} & $2.037\pm0.003$ & 4.29$^{a}$ & \multirow{4}{*}{704055010}\\
			   & XIS-BI & & & $2.534\pm0.004$ & 4.35$^{a}$ & \\
			   & HXD/PIN & & 103800 & $0.135\pm0.003$ & 5.48$^{b}$ & \\
			   & HXD/GSO & & 89228 & $<0.3$ & $<4.80^{c}$ & \\
			\hline
		Swift  & BAT & -- & -- & ($14.5\pm0.3)\times10^{-4}$ & 6.63$^{d}$ & N/A \\
			\hline
		\end{tabular} \\
	$^{a}$ 2-10\,keV flux. \\
	$^{b}$ 15-50\,keV flux. \\
	$^{c}$ 50-100\,keV flux 90\% upper limit \\
	$^{d}$ 20-100\,keV flux. 
\label{observations}
\end{minipage}
\end{center}
\end{table*}

\section{The Suzaku \& Swift Observations of MR\,2251-178}
\label{data-reduction}
\textit{Suzaku} (Mitsuda et al. 2007) observed MR\,2251-178 in the XIS nominal pointing position for 287\,ks between the $7-10^{\rm th}$ of May $2009$. Data are included from the X-ray Imaging Spectrometer (XIS; Koyama et al. 2007) and from the PIN instrument of the Hard X-ray Detector (HXD; Takahashi 2007), both of which were processed using version 2.3.12.25 of the \textit{Suzaku} data reduction pipeline.

A complementary hard X-ray dataset obtained with the \textit{Swift} Burst Alert Telescope (BAT), which observed MR\,2251-178 as part of the 58 month all sky survey (Baumgartner et al. 2010), is also included in our analysis. A summary of the observations is included in Table \ref{observations}.

\subsection{XIS Data Reduction} 
\label{xis-data-analysis}
XIS data were selected in the $3 \times 3$ and $5 \times 5$ modes using ASCA grades 0,\,2,\,3,\,4\,and\,6. Standard XIS screening criteria were applied such that data were excluded if taken: (1) within 436\,s of passage through the South Atlantic Anomaly (SAA), (2) within an Earth elevation angle (ELV) $<\,5^{\circ}$, and/or (3) with Earth day-time elevation angles $<\,20^{\circ}$. Hot and flickering pixels were removed from the XIS images using the \textsc{cleansis} script. Source spectra were extracted from within circular regions of radius $2.8^\prime$, and background spectra were extracted from offset annuli of the same radius with care taken to avoid the corners containing the Fe$^{55}$ calibration sources. The redistribution matrix file and ancillary response file were generated using the tasks \textsc{xisrmfgen} and \textsc{xissimarfgen}, respectively. 

After checking for consistency (see \S \ref{initial-continuum-modelling}), spectra obtained from the two front-illuminated XIS\,0 and XIS\,3 detectors were combined into a single source spectrum (hereafter referred to as XIS-FI) using \textsc{mathpha} in order to maximise signal-to-noise. Data from the back-illuminated XIS\,1 (hereafter XIS-BI) were not combined and were instead analysed separately. Only XIS data over the energy range 0.6--9.0\,keV were included due to a degradation in S/N above 9.0\,keV. Data were also ignored between 1.6--2.1\,keV due to uncertainties associated with the the Si\,K edges intrinsic to the detector assembly of the XIS instrument.

The background subtracted count rates during the observation were $2.037\pm0.003\,\rm{s}^{-1}$ per CCD for the XIS-FI and $2.534\pm0.004\,\rm{s}^{-1}$ in the case of the XIS-BI. The net exposure for the observation was 136.9\,ks.
All XIS spectra and corresponding response files were binned to sample the half-width half-maximum (HWHM) energy resolution of the detectors (i.e.\ $\sim$60\,eV resolution at 6\,keV). Counts were additionally grouped with \textsc{grppha} to achieve a minimum of 50 counts per energy bin to enable the use of $\chi^{2}$ minimisation, which was used for all subsequent spectral fitting.

\subsection{HXD/PIN \& GSO Data Reduction}
\label{hxd-data-analysis}
The HXD/PIN spectrum was extracted from the cleaned events files and was also processed according to the screening criteria described previously. The Non X-ray Background (NXB) was generated using the tuned event files made available by the HXD instrument team (Fukazawa et al. 2009) with a count rate of $0.461\pm0.001$\,counts\,s$^{-1}$. The Cosmic X-ray Background (CXB) was simulated using the form of Boldt (1987) and resulted in a CXB count rate of $\sim0.024$\,counts\,s$^{-1}$. Simultaneous good time intervals (GTIs) were found for both the source spectrum and the NXB using the \textsc{mgtime} task and the NXB exposure time was increased by a factor of ten to reduce the effects of photon noise. The NXB and CXB were combined with equal weight using the \textsc{mathpha} operation and subsequently subtracted from the source spectrum within \textsc{xspec}.

The background subtracted count rate in the HXD/PIN was $0.135\pm0.003$\,counts\,s$^{-1}$, corresponding to a 15-50\,keV flux of $5.48\times10^{-11}$\,erg\,cm$^{-2}$\,s$^{-1}$. The net exposure time was 103.8\,ks after detector dead time was taken into account with \textsc{hxddtcor}. Spectra were binned to a 5$\sigma$ level above the background per bin from 15-50\,keV.

For consistency we also reduced the data obtained with the HXD/GSO. The total (background subtracted) GSO exposure was 89.2\,ks for a count rate of $<0.3$\,counts\,s$^{-1}$, which corresponds to 90\% upper limit on the 50-100\,keV flux of $<4.8\times10^{-11}$\,erg\,cm\,s$^{-1}$. Note that the GSO data are not used during spectral fitting and are used solely to check the consistency of the time-averaged {\it Swift}/BAT dataset.

\section{Broad-band X-ray Spectral Analysis}
\label{preliminary-fitting}
\begin{figure}
\rotatebox{-90}{\includegraphics[height=8.5cm]{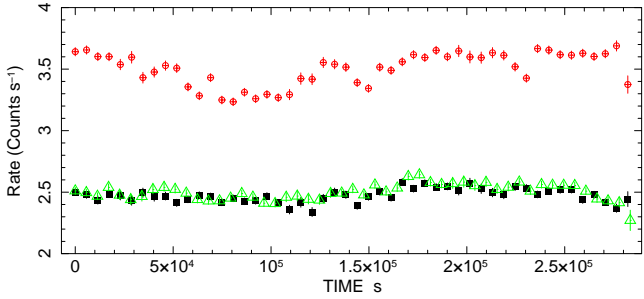}}
\caption{Lightcurves for the XIS0 (green triangles), XIS1 (red circles) and the XIS3 (black closed squares). Only small variations in the lightcurve were observed during the observations. 1$\sigma$ error bars are roughly the size of the plot points.}    
\label{fig:lightcurve}
\end{figure}

We performed a wide-band spectral analysis of MR\,2251-178 in the 0.6--180\,keV energy range using \textsc{xspec} v.\,$12.6.0q$ (Arnaud 1996) and \textsc{heasoft} v.\,$6.10$. All fits were modified by a Galactic column of $N_{\rm H(Gal)}=2.42\times 10^{20}\,\rm{cm}^{-2}$ (Dickey \& Lockman 1990) which was not allowed to vary. We used a photoelectric absorber with Wisconsin cross-sections (the {\tt wabs} component in {\sc xspec}; Morrison \& McCammon 1983) to model the Galactic absorption. Spectral parameters are quoted in the rest-frame of the source ($z=0.064$) unless otherwise stated and errors are quoted to the 90\% level for 1 parameter of interest (i.e.\ $\Delta \chi^{2}=2.71$). The cross-normalisations for the XIS-FI and XIS-BI were allowed to vary in all fits and generally found to be within $\pm2-3\%$ of each other. The HXD/PIN normalisation was set to 1.16 of the XIS. The normalisation on the \textit{Swift}/BAT was left as a free parameter throughout, and was found to be 4-5\% greater than that of the XIS instruments. There was no strong variability (i.e. $<10\%$ level) observed over the course of the observation. Lightcurves for the three XIS detectors are shown in Fig \ref{fig:lightcurve}. In all spectral plots XIS data are binned to sample the half-width half-maximum energy resolution of the detectors (i.e. 60\,eV at 6\,keV) unless otherwise stated.

\subsection{Initial Continuum Parameterisation}
\label{initial-continuum-modelling}
We first considered the X-ray spectrum of MR\,2251-178 in the 3--5\,keV band of the XIS data. 
An initial fit of the continuum with a simple power--law of photon index $\Gamma=1.56\pm0.01$ modified solely by Galactic absorption yielded a reasonable fit (with $\chi^{2}$/\rm{degrees of freedom}=212.5/178). 
Individually, the photon indices for the three XIS spectra are all in good agreement with this value, and have indices of $\Gamma = 1.55\pm0.01$, $\Gamma = 1.55\pm0.02$ and $\Gamma = 1.57\pm0.02$ for the XIS\,0, XIS\,1 and XIS\,3 data, respectively. 

\begin{figure*}
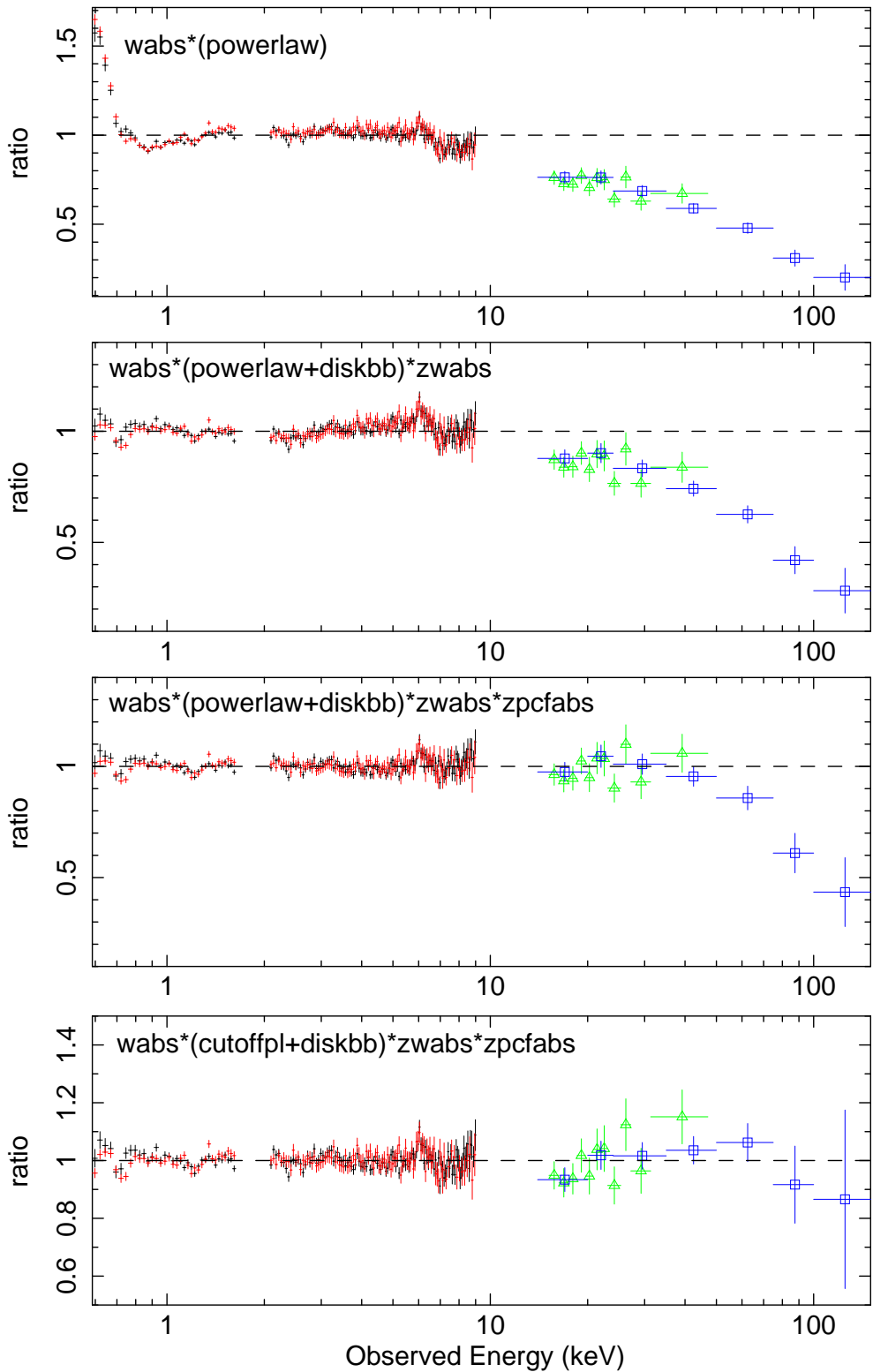

\rotatebox{-90}{\includegraphics[height=13cm]{fig1_noabs_new.ps}}
\rotatebox{-90}{\includegraphics[height=13cm]{fig1_monoabs_new.ps}}
\rotatebox{-90}{\includegraphics[height=13cm]{fig1_dualabs_new.ps}}
\rotatebox{-90}{\includegraphics[height=13cm]{fig1_cutoff.ps}}
\caption{Broadband data/model residuals of the \textit{Suzaku} (XIS-FI: black, XIS-BI: red, PIN: green open triangles) and \textit{Swift} (BAT: blue open squares) data when fit with a variety of parameterising models (\S \ref{initial-continuum-modelling}). From top to bottom: (1) Residuals when fit solely by the 3--5\,keV best-fit power--law (2) Residuals of the same data when fit an accretion disk black body and power-law absorbed by a fully-covering neutral absorber. (3) As above, except with an additional partial-covering absorber included in the model. (4) As above, but with a cut-off power-law. Note that the Fe\,K$\alpha$ has not been fit for illustrative purposes.}
\label{fig:continuumratio}
\end{figure*}

Extending the data to include the full 0.6--180.0\,keV energy range reveals significant deviations from the simple power--law fit and the fitting statistic is extremely poor ($\chi^{2}/\rm{dof}=4733.3/348$). 
The residuals of this fit are shown in Fig. \ref{fig:continuumratio} (top panel). 
In the XIS data a clear soft excess can be seen at energies below around 0.7\,keV and there is a noticeable discrepancy between 0.8--1.6\,keV, presumably due to the presence of the warm absorber. The ubiquitous Fe\,K$\alpha$ line is present at $\sim 6.4$\,keV and the spectrum also appears to roll over above 10\,keV in accordance with the presence of the high-energy cut-off reported by Orr et al. (2001). Note that as the {\it Swift}/BAT spectrum is time-averaged over 58 months, we checked the HXD/GSO spectrum to determine the validity of the high-energy cut-off present in the BAT data. We find a $90\%$ upper limit for the HXD/GSO flux of $<4.8\times10^{-11}$\,erg\,cm$^{-2}$\,s$^{-1}$ in the 50-100\,keV band, which is consistent with the {\it Swift}/BAT flux over the same energy range ($2.7\times10^{-11}$\,erg\,cm$^{-2}$\,s$^{-1}$) and suggests that the roll-over is valid for the {\it Suzaku} observation.

To parameterise the 0.6--180\,keV continuum we initially fit the data with a model of the form $\rm {\tt wabs} \times (\rm {\tt powerlaw} + \rm {\tt diskbb} + \rm {\tt zgauss}) \times \rm {\tt zwabs}$, where the {\tt diskbb} component is a simple parameterisation of the soft excess as an accretion disk blackbody, {\tt zgauss} is a Gaussian used to model the Fe\,K$\alpha$ fluorescence emission line and {\tt zwabs} is a simple photoelectric absorber modelled in the source rest-frame (i.e. at z=0.064). 

With this model the fit statistic is drastically improved, but is still quite poor ($\chi^{2}/\rm{dof}=1111.1/340$). The power-law component has a photon index of $\Gamma=1.67\pm0.01$ and is absorbed by $N_{\rm H}=1.1\times10^{21}$\,cm$^{-2}$, which is similar to that found by Ram\`irez et al. (2008) for the lowly ionised absorber in the {\it Chandra}/LETG observation of this source. The accretion disk blackbody used to parameterise the soft-excess has a temperature $kT=62^{+1}_{-3}$\,eV. The Fe\,K$\alpha$ line is unresolved ($\sigma<391$\,eV) and found at a rest-frame energy of $6.43^{+0.04}_{-0.03}$\,keV, which is consistent with an origin in neutral or mildly ionised material. As shown in Figure \ref{fig:continuumratio} (second panel), the poor fit statistic is due to the model being unable to adequately reproduce the spectral curvature between $\sim2-6$\,keV and in the hard X-ray band. This suggests that additional absorption is required. 

We therefore tested for a more complex absorption scenario by adding an additional neutral absorber with $N_{\rm H}=8.7^{+0.6}_{-0.7}\times10^{22}$\,cm$^{-2}$ which covers $25\pm1\%$ of the source flux (modelled with {\tt zpcfabs} in {\sc xspec}). This gives a significant improvement to the fit ($\chi^{2}/\rm{dof}=565.5/338$ for two additional free parameters) and the residual spectral curvature in the XIS data is no longer present. With a softer photon index of $\Gamma=1.83\pm0.01$, the new model was also able to replicate the observed flux in the HXD/PIN data. Nevertheless, visual inspection of the third panel in Figure \ref{fig:continuumratio} shows that the model is still unable to reproduce the curvature present above $\sim 50$\,keV (Figure \ref{fig:continuumratio}, third panel).

In order to fit the observed rollover in the {\it Swift}/BAT data we replaced the power-law with a {\tt cutoffpl} component which models a power-law with an exponential high-energy cut-off ($E_{\rm cut}$). The addition of the cut-off further improves the fit by $\Delta\chi^{2}=63.2$ for 1 more free parameter and there are no longer any residuals in the BAT data (Figure \ref{fig:continuumratio}, bottom panel). In this model $\Gamma=1.72\pm0.03$ and $E_{\rm cut}=116^{+31}_{-21}$\,keV, {\tt zwabs} has $N_{\rm H}=(1.1\pm0.1)\times10^{21}$\,cm$^{-2}$, {\tt zpcfabs} covers $17\pm2\%$ of the source flux with $N_{\rm H}=9.9^{+1.3}_{-1.2}\times10^{22}$\,cm$^{-2}$ and the soft excess is parameterised by a $kT=57\pm2$\,eV accretion disk blackbody. The final fit statistic is $\chi^{2}/\rm{dof}=502.3/337$ and is our best-fit continuum parameterisation.
 
\subsection{Absorption Lines}
\label{absorption-lines}
Several absorption features are present below 3\,keV and in the Fe\,K band. We modelled these absorption lines with Gaussian profiles fit in the rest-frame of the AGN. The statistical significances for each absorption line in this section were determined against the best-fit continuum model discussed above. Monte Carlo simulations (MC) were also performed to further assess the statistical significances; the details of which are discussed in \S \ref{montecarlo}. Table \ref{table:lines} shows a summary of all line parameters. Detailed discussion regarding all line identifications, including the consideration of alternative identifications, is presented in appendix A.

\subsubsection{The Soft X-ray band}
\label{soft-xray-absorption}
\begin{figure}
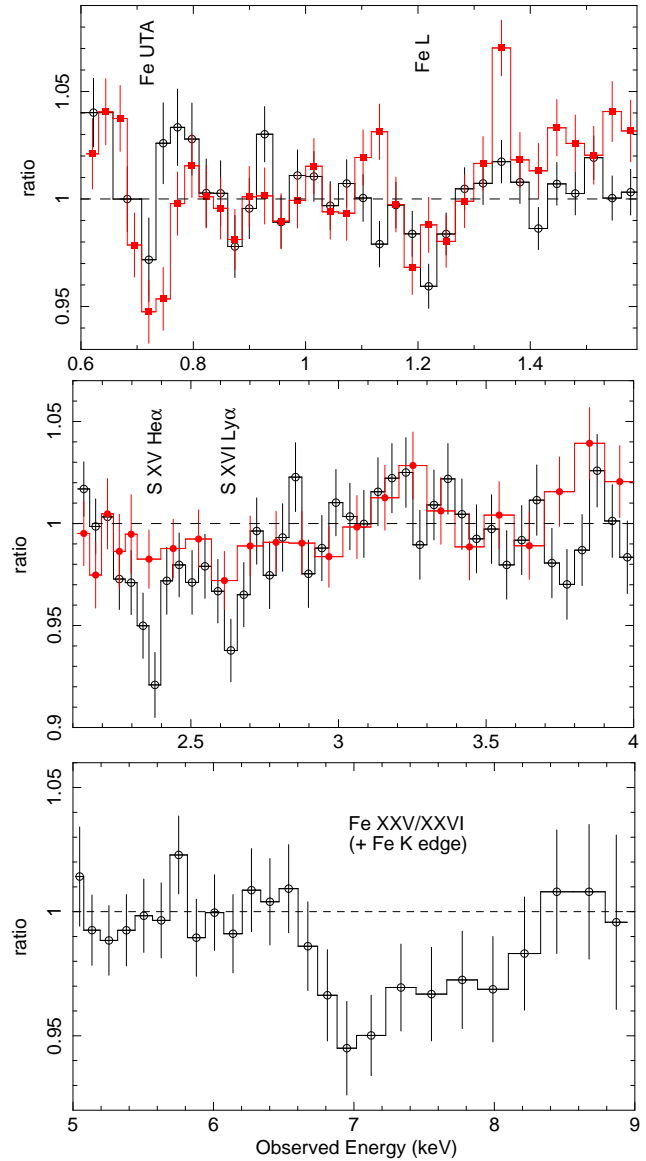

	\begin{center}
					
	    \rotatebox{-90}{\includegraphics[height=8.2cm]{uta_fe24.ps}}
		\rotatebox{-90}{\includegraphics[height=8.5cm]{sulphur.ps}}
		\rotatebox{-90}{\includegraphics[height=8.5cm]{fe_k.ps}}
	\end{center}
	\caption{XIS-FI (black open circles) and XIS-BI (red closed circles) residuals when fit with the best-fit continuum parameterisation, as discussed in \S \ref{initial-continuum-modelling}. The positions of the absorption lines are clearly labelled. The XIS-BI data have been omitted in the bottom panel due to reduced S/N at Fe\,K and the remaining XIS-FI data have been additionally binned by a factor of 3 above the HWHM energy resolution of the detectors for clarity.} 
\label{fig:lineratio}
\end{figure}

Below 2\,keV two absorption lines are required (Fig. \ref{fig:lineratio}; top panel). The first is unresolved and detected at a rest-frame energy of $E=0.77\pm0.01$\,keV. This is consistent with the Fe\,\textsc{i-xvi} M-shell unresolved transition array (UTA) which is expected between 0.729--0.775\,keV (Behar et al. 2001, Netzer et al. 2004). The line has an equivalent width (EW) of $-7\pm2$\,eV and its addition to the model improves the global fit by $\Delta \chi^{2}=65.6$. The line is found to be $>$99.9\% significant from MC simulations. 

The second line has a rest energy of $1.29\pm0.01$\,keV and an EW=$-4^{+2}_{-1}$\,eV.  The line is highly significant ($>99.9\%$ from MC simulations) and its addition improves the fit by $\Delta \chi^{2}=66.1$. Unlike the UTA, this second line does not have an energy that corresponds to any expected strong X-ray transitions. There are several possible identifications for this line (see Appendix A) but the most conservative association which is internally consistent with our {\sc xstar} model (\S\ref{xstar-modelling}) is with the Fe\,\textsc{xxiv} 2s$\rightarrow$3p doublet which is expected at an mean energy of $E=1.165$\,keV. The line identification requires a blueshift of $v_{\rm out}=0.11\pm0.01$\,c ($33000\pm3000\,\rm{km\,s}^{-1}$) assuming this identification.

Two further Gaussian absorption lines are required at $E=2.52\pm0.02$\,keV and $E=2.79\pm0.03$\,keV in the rest-frame, with equivalent widths of $-6\pm2$\,eV and $-5\pm2$\,eV, respectively (Fig. \ref{fig:lineratio}; middle panel). Both lines are significant at the $>99.9\%$ level from MC simulations and their addition improves the model by a further $\Delta\chi^{2}=15.9$ and $\Delta\chi^{2}=10.9$, respectively. Theoretically, the strongest spectral features expected in the 2--3\,keV energy range are those associated with the 1s$\rightarrow$2p transitions of S\,\textsc{xv} ($E=2.461$\,keV) and S\,\textsc{xvi} ($E=2.623$\,keV), but the rest-frame energies of the detected lines do not correspond to either of these which suggesting that they may be blueshifted. If identified with the S\,\textsc{xv} and S\,\textsc{xvi} transitions the measured line centroids indicate that they are blueshifted by $v_{\rm out}=0.02\pm0.01$\,c ($6000\pm3000\,\rm{km\,s}^{-1}$) and $v_{\rm out}=0.07\pm0.01$\,c ($21000\pm3000\,\rm{km\,s}^{-1}$), respectively. Both lines are unresolved and their widths were fixed to $\sigma=10$\,eV throughout. 

\subsubsection{The Fe\,K region}
\label{feK-absorption}
There are further residuals in the data/model ratio present in the Fe\,K region. In addition to the Fe\,K$\alpha$ emission line there is a clear broad absorption feature in the 6.5--8\,keV range, as can be seen in the bottom panel of Fig. \ref{fig:lineratio}. Note that this absorption is found in excess of the neutral Fe\,K absorption edge (expected at E=7.1\,keV) which is included as part of both the {\tt zwabs} and {\tt zpcfabs} absorbers. This absorption trough was initially modelled with a broad Gaussian profile and its centroid energy left as a free parameter. The centroid energy and intrinsic width of the broad line are $E=7.57^{+0.19}_{-0.12}$\,keV and $\sigma=209^{+244}_{-108}$\,eV, and it has an EW of $-26^{+18}_{-12}$\,eV. The addition of the line results in a $\Delta\chi^{2}$ improvement of $16.6$ for 3 additional free parameters, and it is statistically required at the 99.3\% level from MC simulations. With the addition of the broad absorption component the width of the Fe\,K$\alpha$ emission line is now more tightly constrained than before at $\sigma<117$\,eV (while EW=$25^{+12}_{-8}$\,eV).

In this energy range the spectrum is expected to be dominated by atomic features attributable to iron. In particular, the 1s$\rightarrow$2p transitions are expected to be particularly strong, as are the many resonance transitions associated with the Fe\,K edge near 7.1\,keV (Kallman et al. 2004). If due to the strong 1s$\rightarrow$2p transitions this immediately places a constraint on the possible identifications for the absorption as being due to at least Fe\,\textsc{xviii} ($E\sim6.5$\,keV), where the Fe ions become sufficiently ionised to have an L-shell vacancy for a 1s electron transition. Above this ionisation state, the 1s$\rightarrow$2p transitions occur between $E\sim6.5-6.6$\,keV for Fe\,\textsc{xviii-xxiv}, and at $E\sim6.7$\,keV and $E\sim6.97$\,keV for Fe\,\textsc{xxv}\,He$\alpha$ and Fe\,\textsc{xxvi}\,Ly$\alpha$, respectively. If the broad profile is identified with Fe\,\textsc{xxv} He$\alpha$ it requires a velocity shift of $v_{\rm out}=0.13^{+0.03}_{-0.02}$\,c (= $39000^{+9000}_{-6000}\,\rm{km\,s}^{-1}$), while if identified with Fe\,\textsc{xxvi} Ly$\alpha$ the velocity shift is slightly lower at $v_{\rm out}=0.10^{+0.03}_{-0.02}$\,c ($30000^{+9000}_{-6000}\,\rm{km\,s}^{-1}$). 

\begin{figure}
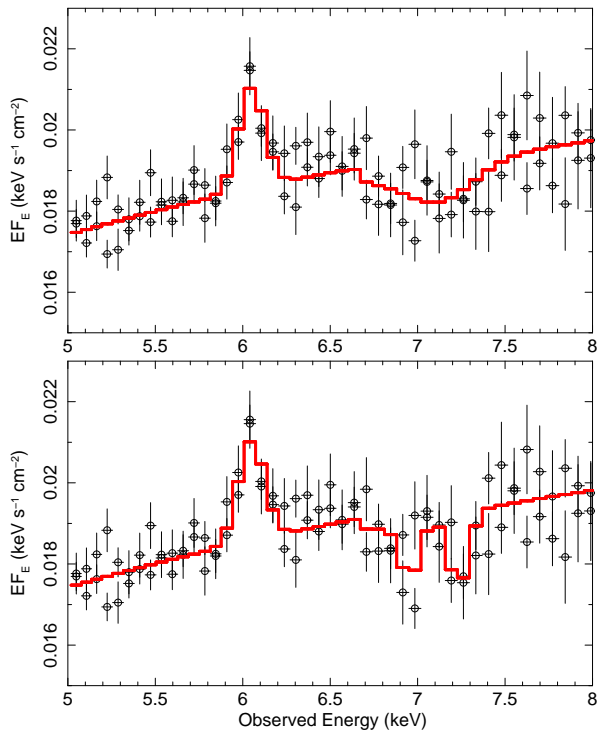

	\begin{center}
		\rotatebox{-90}{\includegraphics[height=8cm]{fe_1line.ps}}
		\rotatebox{-90}{\includegraphics[height=8cm]{fe_2line.ps}}
	\end{center}
	\caption{Plots showing the Fe\,K band. \textbf{Top:} the absorption feature fit with a broad Gaussian. \textbf{Middle:} the same absorption feature when fit as two separate Gaussian lines using the method outlined in \S \ref{feK-absorption}.}
\label{fig:fekgauss}
\end{figure}

An alternative interpretation is that the broad absorption represents a blend of the above lines rather than a single discrete profile. To investigate the possibility that the profile is a blend of He- and H- like Fe we replaced the broad profile with two narrow ($\sigma=50$\,eV) absorption lines at fixed rest frame energies of $6.7$\,keV and $6.97$\,keV and left their common velocity shift as a free parameter. This fit was a slight improvement to that of the broad Gaussian obtained previously, and gives a $\Delta\chi^{2}=18.1$ for 3 parameters of interest. Even so Fig. \ref{fig:fekgauss} shows that this model slightly under-predicts the absorbed flux of the Fe\,\textsc{xxv} He$\alpha$ line at $E=7.36^{+0.09}_{-0.11}$\,keV (rest-frame; $E=6.92$ observed), while the Fe\,\textsc{xxvi} Ly$\alpha$ line is well fit. The resulting simultaneous velocity shift is $v=0.10\pm{0.01}$\,c with respect to the rest frame of the AGN. 

\begin{table*}
\begin{minipage}{150mm}
\caption{Summary of X-ray absorption and emission lines}
\label{table:lines}
\begin{tabular}{@{}cccccccc}
\hline
Line ID & $E_{\rm rest}$\,(keV) & $E_{\rm lab}$\,(keV) & Flux$^{a}$ & $\sigma$\,(eV) & EW\,(eV) & $\Delta\chi^{2\,b}$ & MC \\ \hline\hline
Fe\,UTA  &  $0.77\pm0.01$ & $0.729-0.775$ & $-241.0^{+67.0}_{-60.0}$ & $10^{*}$ & $-7\pm2$ & 65.6 & $>99.9\%$ \\ [0.5ex]

Fe\,{\sc xxiv} & \multirow{2}{*}{$1.29\pm0.01$} & \multirow{2}{*}{$1.165$} & \multirow{2}{*}{$-38.0^{+14.7}_{-11.3}$} & \multirow{2}{*}{$<44$} & \multirow{2}{*}{$-4^{+2}_{-1}$} & \multirow{2}{*}{66.1} & \multirow{2}{*}{$>99.9\%$} \\ 
(2s$\rightarrow$3p) & & & & & & \\[1ex]

S\,\textsc{xv} & $2.52\pm0.02$ & 2.461 & $-17.2\pm5.7$ & $10^{*}$ & $-6\pm2$  & 15.9 & $>99.9\% $\\ [0.5ex]

S\,\textsc{xvi} & $2.79\pm0.03$ & 2.623 & $-11.5^{+4.7}_{-4.8}$ & $10^{*}$ & $-5\pm2$ & 10.9 & $>99.9\%$ \\ [0.5ex]

Fe\,K$\alpha$ & $6.43^{+0.04}_{-0.03}$ & 6.4 & $+14.0^{+6.7}_{-4.2}$ & $<117 $ & $+25^{+12}_{-8}$ & 52.9 & --- \\ [0.5ex]

Fe\,\textsc{xxv-xxvi} & $7.57^{+0.2}_{-0.1}$ & 6.7 and/or 6.97 & $-8.9^{+6.0}_{-4.7}$ & $183^{+185}_{-112}$ & $-22^{+15}_{-11}$ & 16.6 & 99.3\% \\[1.0ex]
\hline
\end{tabular} \\
$^{*}$ Denotes parameter was fixed at listed value \\
$^{a}$ Line flux quoted in units of $\times10^{-6}$\,erg\,cm$^{-2}$\,s$^{-1}$. \\
$^{b}$ Change in fitting statistic when adding a Gaussian to best-fit continuum model.
\end{minipage}
\end{table*}

\subsection{Montecarlo Simulations}
\label{montecarlo}
The Montecarlo simulations we used are analogous to those carried out by Tombesi et al. (2010a). Briefly: 

\begin{enumerate}
\item We simulated $S=1000$ XIS spectra between 0.6-10.0\,keV based upon the baseline continuum parameterisation model (outlined in \S \ref{initial-continuum-modelling}), using the \textit{fakeit} command in \textsc{xspec}. Both the HXD/PIN and \textit{Swift}/BAT spectra were not included in the simulation as they had no effect on the absorption line parameters. 
\item In each of the simulated spectra we stepped an inverted Gaussian of a set fixed width (see below) between two pre-defined energy bands in equal steps. After each step, the model was fit for the line normalisation of the Gaussian and the resulting overall $\Delta\chi^{2}$ was recorded. To sample the entire XIS spectrum we performed this process twice, once to cover the Fe\,K band and once to  cover the soft X-rays. For the Fe\,K absorption we stepped a broad absorption line of $\sigma=200$\,eV width every 100\,eV between 4-9\,keV, while in the soft X-ray band we stepped a narrow absorption line of 10\,eV width every 25\,eV between 0.6 and 4\,keV. Due to the XIS-BI having poorer S/N around Fe\,K we used only the XIS-FI data in this band. Both the XIS-FI and XIS-BI were used in the soft X-rays where the S/N of the XIS-BI is higher.
\item After $S$ spectra were generated the associated grid of $\chi^{2}$ values then corresponded to how statistically likely it would be for a randomly occurring feature to have a $\Delta\chi^{2}$ improvement greater than a particular value. To assess the probabilities of the absorption lines detected, the $\Delta\chi^{2}$ value we obtained from our Gaussian fits was then compared to this generated grid of values. If $N$ of the simulated spectra had a feature of significance greater than this $\Delta\chi^{2}$ value, the resulting detection confidence level of the measured line was $1-N/S$. 
\end{enumerate}

The statistical probabilities for each line from MC simulations are recorded in Table \ref{table:lines}.

\section{Self-consistent Modelling}
\label{xstar-modelling}
With both the general continuum and the absorption lines parameterised, we proceeded to replace the simple parameterising models (i.e.\ \texttt{zpcfabs} and \texttt{zwabs}) with a series of self-consistent model grids generated using the \textsc{xstar} photoionisation code. Each \textsc{xstar} grid contains a series of models, each pertaining to a particular photoionised spectrum with a characteristic ionisation parameter ($\xi$), a column density ($N_{\rm H}$), and a redshift ($z$).  In addition, each grid is attributed an intrinsic turbulent velocity ($V_{\rm turb}$) upon generation which determines the intrinsic width of any emission and absorption lines. The grid is then used as a simple multiplicative model in {\sc xspec} to simultaneously fit the broadband continuum and any spectral lines which may be present.

\subsection{Fully covering models}
We first modelled the spectrum assuming that the photoionised absorbers fully cover the view to the ionising source, and that the soft excess represents an intrinsic component of the source continuum. We sequentially added \textsc{xstar} grids to a baseline continuum model of the form \texttt{wabs} $\times$ (\texttt{cutoffpl+ diskbb+zgauss}) $\times$ \texttt{xstar} until an acceptable fit had been achieved. Solar abundances of Grevesse \& Sauval (1998) are assumed throughout unless otherwise stated.

\subsubsection{Lowly ionised absorbers} 
\begin{figure}
	\begin{center}
		\rotatebox{-90}{\includegraphics[height=8cm]{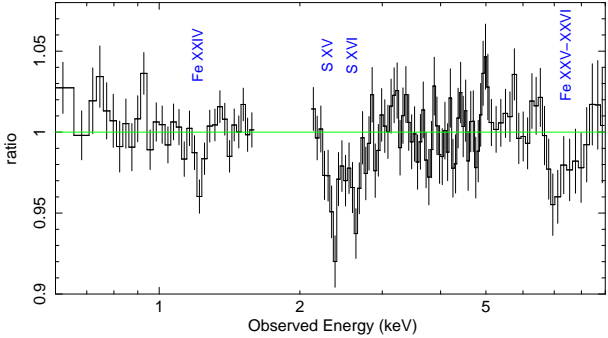}}
	\end{center}
	\caption{Data/model residuals of the XIS-FI data when fitted with two lowly ionised soft X-ray absorbers. The position of the Fe\,L, S\,{\sc xv}, S\,{\sc xvi} and Fe\,K absorption is clearly apparent. For clarity, the XIS-BI data have again been omitted. Data between 1.6-2.1\,keV have been removed due to the instrumental Si\,K edge. See text for further details.}
\label{fig:2sratio}
\end{figure}

In the soft X-ray band two low turbulence ($v_{\rm turb}=200$\,km\,s$^{-1}$) ionised absorption components are required to provide a good description to both the Fe UTA absorption line and the general spectral curvature below 10\,keV (see Fig. \ref{fig:2sratio}).
The first absorber (Zone 1; $\Delta\chi^{2}=93.7$ for 3 additional free parameters) is lowly ionised and mainly responsible for the Fe\,UTA, while he second absorber (Zone 2) is less significant to the fit ($\Delta\chi^{2}=36.9$ for 2 more free parameters) and mainly fits the curvature. These absorbers also contribute to a shallow Fe\,K edge in the Fe\,K band, however its depth is insufficient to adequately replicate the feature (see \S\ref{highly-ionised}). For fitting purposes the soft X-ray absorbers were assumed to be co-spatial and their common outflow velocity was fixed to $v_{\rm out}=300$\,km\,s$^{-1}$ to be consistent with the UV outflows reported in this source by Monier et al. (2001). Allowing the outflow velocities to vary did not yield any appreciable change in the fit statistic and they were thus left fixed throughout.

\subsubsection{Detection of a Fe\,L-shell and Fe\,\textsc{xxv} absorber?}
\label{fel_abs}
\begin{figure}
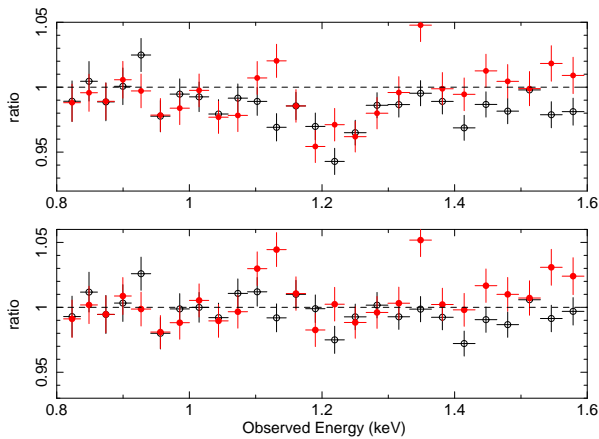

  \begin{center}
    \rotatebox{-90}{\includegraphics[height=8cm]{ratio_fe24_noabs.ps}}
    \rotatebox{-90}{\includegraphics[height=8cm]{ratio_fe24_v5000.ps}}
  \end{center}
  \caption{Ratio plots of the 0.8--1.6\,keV energy band after the inclusion of absorption zones 1 and 2, showing: \textbf{Top:} the residual Fe\,L line profile. \textbf{Bottom:} the same line but when zone 3 of the photoionisation model is added. The residuals are clearly no longer present.}
\label{fig:fe24ratio}
\end{figure}

\begin{figure}
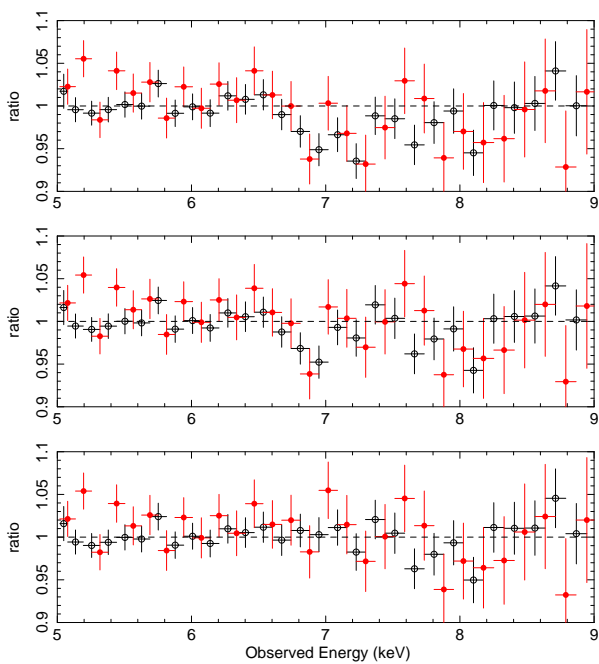

  \begin{center}
    \rotatebox{-90}{\includegraphics[height=8cm]{ratio_5-9keV_noabs.ps}}
    \rotatebox{-90}{\includegraphics[height=8cm]{ratio_5-9keV_1abs.ps}}
    \rotatebox{-90}{\includegraphics[height=8cm]{ratio_5-9keV_2abs.ps}}
    \end{center}
  \caption{Ratio plots of the Fe\,K band in the fully-covering \textsc{xstar} model. \textbf{Top:} when no ionised absorbers are included two residual profiles are apparent between 6.5--8.0\,keV. \textbf{Middle:} when Zone 3 is added the higher energy residual is no longer present. \textbf{Bottom:} when a further highly-ionised zone is added there are no remaining residuals. See text for further details.}
\label{fig:fekratio}
\end{figure}

To model the absorption line at $\sim1.29$\,keV we used a higher turbulence ($v_{\rm turb}=5000$\,km\,s$^{-1}$) grid to account for the apparent line broadening. Zone 3 ($\Delta\chi^{2}=74.0$; see Table\ref{final_fcov} for parameters) fits both the 1.29\,keV line (Fig \ref{fig:fe24ratio}) and also appears to fit the higher energy component of the Fe\,K absorption (Fig. \ref{fig:fekratio}) at $E\sim7.7$\,keV (rest-frame; 7.24\,keV observed). The outflow velocity found for this absorber is consistent with an identification of the Fe\,K absorption with Fe\,\textsc{xxv} He$\alpha$, however, the velocity is substantially larger than that found if the 1.29\,keV line is identified as solely Fe\,\textsc{xxiv}. This could be the result of the line at 1.29\,keV being a blend of Fe\,{\sc xxiii} (2s$\rightarrow$3p; expected at 1.127\,keV) and Fe\,{\sc xxiv} (2s$\rightarrow$3p). If this is the case, the outflow velocity would then be consistent with that found for the {\sc xstar} grid. We hereafter denote this absorption component the ``Fe\,L-shell absorber'' as at an ionisation of $\log\xi/\rm{erg\,cm\,s}^{-1}=3.02$, the most prominent ions are from Fe\,{\sc xxiii-xxv}.

\subsubsection{Sulphur absorbers}
\label{sulphur_xstar}
There is a level of ambiguity regarding the correct physical interpretation of the lines at $2.52$ and $2.79$\,keV. In particular, we were unable to obtain a satisfactory fit to the lines assuming solar abundances, and the lack of other distinct absorption lines meant that when fit with variable abundances, the required Sulphur over-abundance was entirely unconstrained. With the current data we therefore parameterise the absorption lines using a column density comprising solely of Sulphur ($N_{\rm S}$).

As discussed in \S \ref{soft-xray-absorption}, the most conservative interpretation is that the lines are identified with the 1s$\rightarrow$2p transitions of He- and H-like Sulphur, respectively. In this interpretation the lines are well modelled by the addition of two further ionised absorbers. The first (Zone 4), which models S\,\textsc{xv} at $\sim2.5$\,keV, is described $\log\xi/\rm{erg\,cm\,s}^{-1}=2.44^{+0.24}_{-0.22}$ and a Sulphur column density of $N_{\rm S}=1.9^{+0.3}_{-0.8}\times10^{17}$\,cm$^{-2}$. For a solar Sulphur/Hydrogen abundance ratio of $2.14\times10^{-5}$ (Grevesse \& Suvaul 1998) this corresponds to a Hydrogen column of $N_{\rm H}=8.9^{+1.4}_{-3.7}\times10^{21}$\,cm$^{-2}$, which is similar to the column densities reported previously for this source (i.e. Orr et al. 2001; Kaspi et al. 2004; Gibson et al. 2005, Ram\`irez et al. 2008). The net outflow velocity of this absorber is $v_{\rm out}=6300^{+5100}_{-4200}$
\,km\,s$^{-1}$ ($0.021^{+0.017}_{-0.014}$\,c), which is consistent with that found when fitting simple Gaussians. At this column density and ionisation parameter the grid also contributes He$\beta$ (1s$\rightarrow$3p) and weak He$\gamma$ (1s$\rightarrow$4p) lines in addition to strong He$\alpha$, however the energy separation of the lines is insufficient to fit the line at $\sim2.8$\,keV simultaneously and a second grid is required to model the remaining residuals.

The second Sulphur absorber (Zone 5; $\Delta\chi^{2}=32.9$) which models the S\,{\sc xvi} Ly$\alpha$ is described by $N_{\rm S}>2.2\times10^{17}$\,cm$^{-2}$ (= $N_{\rm H}>1\times10^{22}$\,cm$^{-2}$) and $\log\xi/\rm{erg\,cm\,s}^{-1}=3.37^{+0.12}_{-0.41}$. The outflow velocity is found to be $v_{\rm out}=21000\pm3000$\,km\,s$^{-1}$ ($0.07\pm0.01$\,c), which is again consistent with that found from simple Gaussian fitting. The fit statistic for the model assuming two zones of Sulphur absorption is $\chi^{2}/\rm{dof}=352.2/316$.

\subsubsection{A further highly ionised zone?}
\label{highly-ionised}
While the soft X-ray and Fe\,L absorbers contribute a shallow Fe\,K edge at $\sim7.1$\,keV (see middle panel of Figure \ref{fig:fekratio}) its depth is insufficient to fit the remaining absorption feature at $E\sim7.36$\,keV (rest-frame, observed frame $E\sim6.92$\,keV). This feature can be tentatively ($\Delta\chi^{2}=8.0$ for 3 additional parameters; Fig \ref{fig:fekratio} bottom panel) fit by the addition of a highly ionised and high column density absorber ($v_{\rm turb}=5000$\,km\,s$^{-1}$) which is outflowing with $v_{\rm out}=18000\pm3000$\,km\,s$^{-1}$ ($0.06\pm0.01$\,c). The parameters imply that the feature may be identified with Fe\,{\sc xxvi} (1s$\rightarrow$2p), which is expected to be the must abundant ion at such an ionisation.  
This feature may be a higher velocity analogue of the outflowing Fe\,{\sc xxvi} absorption ($v_{\rm out}\sim12700$\,km\,s$^{-1}$) which was reported by Gibson et al. (2005) in a earlier {\it Chandra}/HETG observation of this source.
 
\subsubsection{Reflection Component}
The above absorbers in addition to the {\tt cutoffpl}, {\tt  diskbb} and {\tt zgauss} components provide a good description of the general broad-band continuum and the fit is marginally acceptable ($\chi^{2}/\rm{dof}=352.2/316$; null probability=0.08). However, for consistency we replaced the redshifted Gaussian modelling the Fe\,K$\alpha$ emission with the ionised reflection model {\tt reflionx} (Ross \& Fabian 2005), which self-consistently takes into account any reprocessing due to reflection in a face-on system. Fixed Solar abundances were assumed and the photon-index of the reflected power-law was tied to that of the intrinsic. 

The addition of a lowly ionised reflection component ($\xi<27$\,erg\,cm\,s$^{-1}$) with a low reflection fraction ($R<0.2$, where $R=\Omega/2\pi$ and $\Omega$ is the solid angle subtended by the reflector) reproduces the Fe\,K$\alpha$ profile and improves the fit in the hard X-ray band, for a final fit statistic of $\chi^{2}/\rm{dof}=342.79/316$ (null prob=0.14),  $\Gamma=1.67^{+0.02}_{-0.03}$ and $E_{\rm cut}=101^{+30}_{-24}$\,keV). The overall best-fit parameters for the fully-covering model are listed in Table \ref{final_fcov}.

\begin{table}[h]
\begin{center}
\caption{Best-fit parameters for the fully covering model}
\begin{tabular}{c c c c}
\hline
\multicolumn{4}{c}{\textbf{Continuum parameters}} \\ \hline\hline
\multirow{2}{*}{Flux (erg\,cm$^{-2}$\,s$^{-1}$)} & $F_{\rm 0.5-2.0}$ & \multicolumn{2}{c}{$3.0\times10^{-11}$} \\
 & $F_{\rm 2.0-10.0}$ & \multicolumn{2}{c}{$4.5\times10^{-11}$} \\
\multirow{2}{*}{Luminosity (erg\,s$^{-1}$)} & $L_{\rm 0.5-2.0}$ & \multicolumn{2}{c}{$2.83\times10^{44}$} \\
 & $L_{\rm 2.0-10.0}$ & \multicolumn{2}{c}{$4.41\times10^{44}$} \\ 
\hline
\multicolumn{4}{c}{\textbf{Best-fit model components}} \\ [0.5ex]
Component & Parameter & Value & $\Delta\chi^{2\,h}$ \\
\hline\hline
Galactic absorption & $N_{\rm H}^{a}$ & $2.42\times10^{20}$ & --- \\ [1ex]
cutoffpl & $\Gamma$ & $1.68\pm0.01$ & --- \\ [0.5ex]
 & $E_{\rm cut}^{b}$ & $101^{+23}_{-17}$ & \\ [1ex]
reflection & Abund & $1^{*}$ & 77.2$^{\,i}$ \\ [0.5ex]
 & $\xi^{e}$ & $<27$ & \\ [0.5ex]
 & R & $<0.2$ &  \\[0.5ex]
 diskbb & $T_{\rm in}^{c}$ & $62^{+5}_{-6}$ & 114.6 \\ [0.5ex]
 & norm & $2.96^{+1.04}_{-1.32}\times10^{6}$ & \\ [1ex]
Soft X-ray absorber 1 & $N_{\rm H}^{a}$ & $(5.4\pm0.4)\times10^{20}$ & 93.7 \\ [0.5ex]
 (Zone 1) & $\log\xi^{e}$ & $-0.23^{+0.10}_{-0.12}$& \\ [0.5ex]
 & $v_{\rm out}^{f}$ & $300$  \\ [0.5ex]
 & $v_{\rm turb}^{g}$ & 200  \\ [1ex] 
Soft X-ray absorber 2 & $N_{\rm H}^{a}$ & $5.6^{+1.3}_{-2.1}\times10^{21}$ & 36.9 \\ [0.5ex]
 (Zone 2) & $\log\xi^{e}$ & $2.21^{+0.08}_{-0.07}$ & \\ [0.5ex]
 & $v_{\rm out}^{f}$ & $300$  \\ [0.5ex] 
 & $v_{\rm turb}^{g}$ & 200  \\ [1ex] 
Fe\,L absorber & $N_{\rm H}^{a}$ & $5.2^{+6.0}_{-0.9}\times10^{21}$ & 74.0 \\ [0.5ex]
(Zone 3) & $\log\xi^{e}$ & $3.02^{+0.08}_{-0.03}$ \\ [0.5ex]
 & $v_{\rm out}^{f}$ & $42000^{+3000}_{-5000}$ \\ [0.5ex]
 & $v_{\rm turb}^{g}$ & 5000  \\ [1ex]
S\,\textsc{xv} absorber & $N_{\rm H}^{a}$ & $8.9^{+1.4}_{-3.7}\times10^{21}$ & 76.4 \\ [0.5ex]
(Zone 3) & $\log\xi^{e}$ & $2.44^{+0.24}_{-0.22}$ & \\ [0.5ex]
 & $v_{\rm out}^{f}$ & $6300^{+5100}_{-4200}$ \\ [0.5ex]
 & $v_{\rm turb}^{g}$ & 1000  \\ [1ex]
S\,\textsc{xvi} absorber & $N_{\rm H}^{a}$ & $>1.0\times10^{22}$ & 32.9 \\ [0.5ex]
(Zone 5) & $\log\xi^{e}$ & $3.37^{+0.12}_{-0.41}$ & \\ [0.5ex]
 & $v_{\rm out}^{f}$ & $19800^{+4200}_{-5100}$ \\ [0.5ex] 
 & $v_{\rm turb}^{g}$ & 1000 \\ [1ex]
\hline
\label{final_fcov}
\end{tabular} \\
\end{center}
$^{a}$ Absorber column density in units of cm$^{-2}$. \\
$^{b}$ Cut-off energy in units of keV. \\
$^{c}$ Accretion  disk black body temperature, in units of eV \\
$^{e}$ ionisation parameter in units of erg\,cm\,s$^{-1}$. \\
$^{f}$ Outflow Velocity of absorber given in units of km\,s$^{-1}$. \\
$^{g}$ Assumed absorber turbulent velocity in units of km\,s$^{-1}$. \\
$^{h}$ Significance of component with respect to the final best-fitting model. \\
$^{i}$ Statistical significance of {\sc reflionx} component is given in comparison to a fit where Fe\,K$\alpha$ is left unmodelled. \\

\end{table}

\subsection{Partial covering models}
We also investigated the possibility that only a fraction $f<1$ of the line-of-sight source flux may be covered by the ionised absorbers. This would then require $\Gamma$ for the power-law continuum to steepen to compensate for the increased absorption, and similar partial covering scenarios have been invoked to self-consistently model hard X-ray excesses in several AGN (e.g. NGC1365, Risaliti et al. 2009; 1H0419-577, Turner et al. 2009; PDS 456, Reeves et al. 2009). Modelling the hard X-ray excess in this way typically requires the intrinsic flux below 10\,keV to be strongly absorbed by Compton thick material (i.e. $N_{\rm H}>10^{24}$\,cm$^{-2}$). However, if the partial coverer is Compton thin, it may be able replicate the roll-over and soft excess. 

We thus constructed a low turbulence ($v_{\rm turb}=200$\,km\,s$^{-1}$ partial covering model using the {\tt zxipcf} model in {\sc xspec}. Our starting model had the form \texttt{wabs} $\times$ (\texttt{zxipcf$\times$powerlaw+reflionx}) $\times$ \texttt{xstar}, where we have removed the {\tt diskbb} component which previously parameterised the soft-excess and reverted to a simple power-law with no high-energy cut-off. The overall best-fit parameters for our partial covering model are listed in Table $\ref{final_pcov}$. Importantly, this model still requires the presence of several fully covering absorbers to account for the Fe\,L/Fe\,K and Sulphur absorption, and these absorbers have best-fit parameters which are consistent with those obtained in the fully covering model. A {\tt reflionx} reflection component was also included for self consistency. In this geometry the partially covering clouds would need to be very compact in size and located close to the ionising source to only obscure a fraction of the emergent continuum. One possibility is that the partial coverer may be associated with a clumpy disk-wind where the covering fraction relates to the filling factor of the denser clouds, while the more ionised absorbers appear to fully cover the source along the line-of-sight. Such scenarios have been effective in explaining the broad-band spectral properties of objects such as MCG\,-6-30-15 (Miller et al. 2008) and 3C\,445 (Reeves et al. 2010), and are consistent with theoretical expectations that disk-winds are inhomogeneous outflows and contain clumps of dense, high column-density, material (e.g. Proga \& Kallman 2004; Sim et al. 2010)

In this case the partial covering model requires a very lowly ionised, high column density absorber which covers approximately 30\% of the source in order to fit the soft-excess and curvature below 10\,keV (Table \ref{final_pcov}). As expected, the power-law is also much steeper ($\Gamma=1.96\pm0.01$) than that usually inferred for this source ($\Gamma=1.6-1.7$). With just the partial coverer included significant residuals remained in the soft X-ray band and two low turbulence ($v_{\rm turb}=200$\,km\,s$^{-2}$) fully-covering absorbers are also required to achieve an adequate fit (Zones 1 and 2 in Table \ref{final_pcov}). 
 
\begin{figure}
	\begin{center}
		\rotatebox{-90}{\includegraphics[height=8cm]{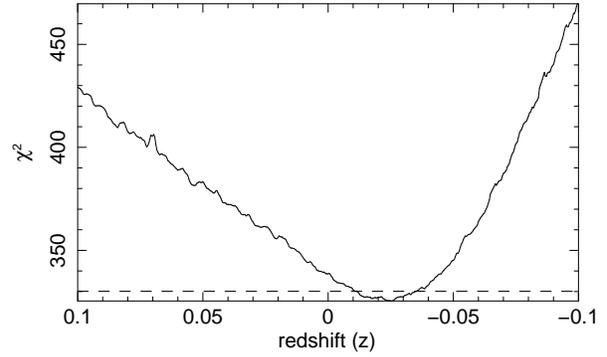}}
	\end{center}
	\caption{Contour plot showing the redshift for the high-velocity soft X-ray absorber in the partial-covering model.}
\label{contour}
\end{figure}

Interestingly, the outflow velocity of Zone 2 is $v_{\rm out}=27000\pm3000$\,km\,s$^{-1}$ ($0.09\pm0.01$\,c), which is much faster than the 100-1000\,km\,s$^{-1}$ expected for typical warm absorbers (e.g. Blustin et al. 2005; Crenshaw, Kraemer \& George 2003). Nevertheless, the very high outflow velocity is strongly required in this model (see Figure \ref{contour}), and fixing the outflow velocity to the source rest-frame (i.e. $v_{\rm out}=0$) 
resulted in a worse fit by $\Delta\chi^{2}=46.6$ (for one less free parameter). 
As it is not required in the fully-covering scenario, the high outflow velocity of this component is likely to be highly model dependent and further observations at a higher resolution are required to determine its validity. 

\subsubsection{The Fe\,K band}
As in the fully-covering model, the {\tt reflionx} component gives an excellent description of the Fe\,K$\alpha$ emission line and the reflector is once again found to be both lowly ionised and have a low reflection fraction. The partial coverer provides a very good fit to the residual absorption in the Fe\,K band (see inset of Figure. \ref{broadband_continuum}, bottom panel) and, in addition to the Fe\,\textsc{xxv} He$\alpha$ absorption which is being modelled by the Fe\,L absorber, the lowly ionised partial coverer provides an excellent fit to the Fe\,K edge.

\subsubsection{The high-energy roll-over}
\begin{figure*}
\includegraphics[width=15cm]{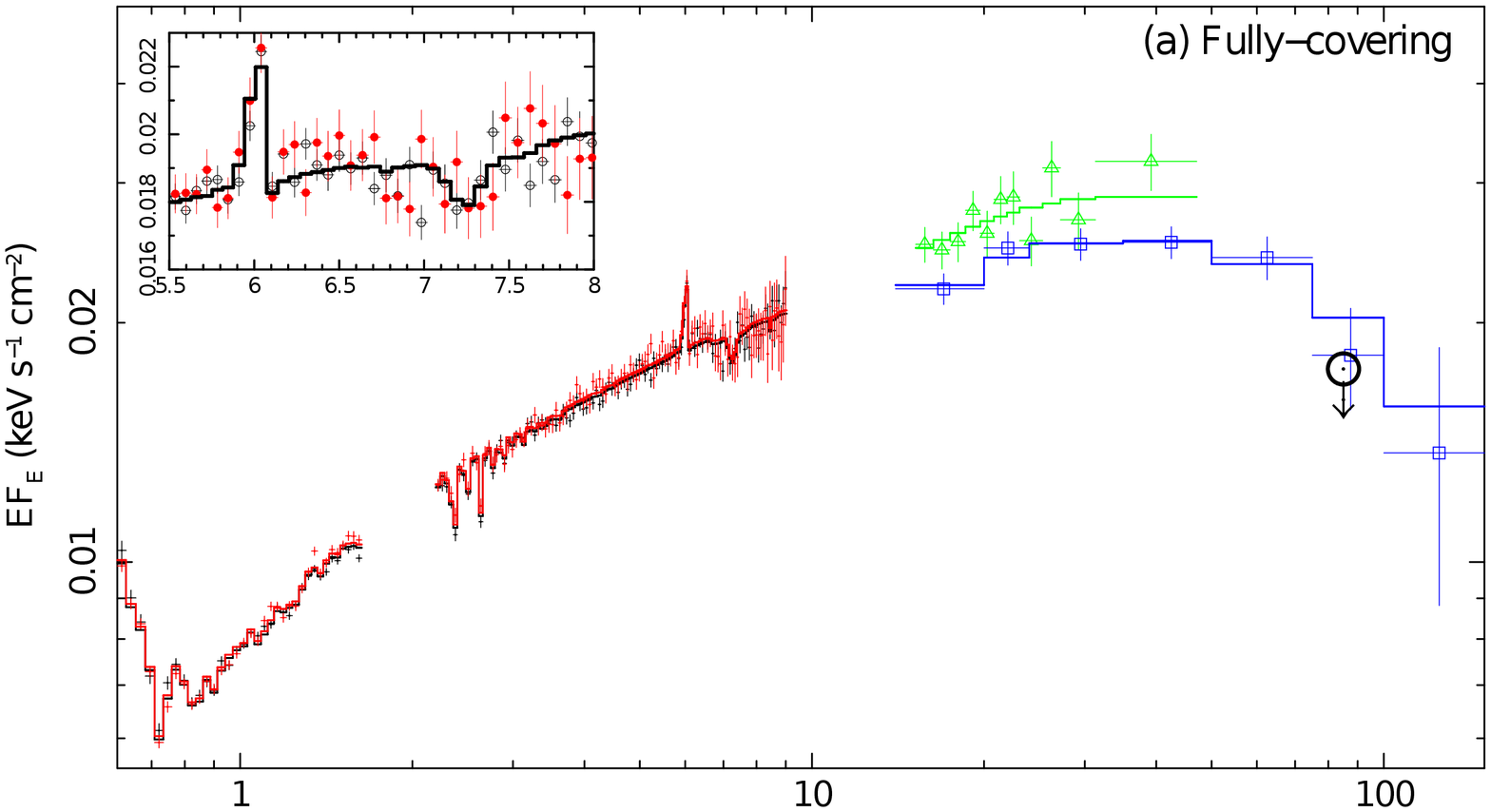}
\includegraphics[width=15cm]{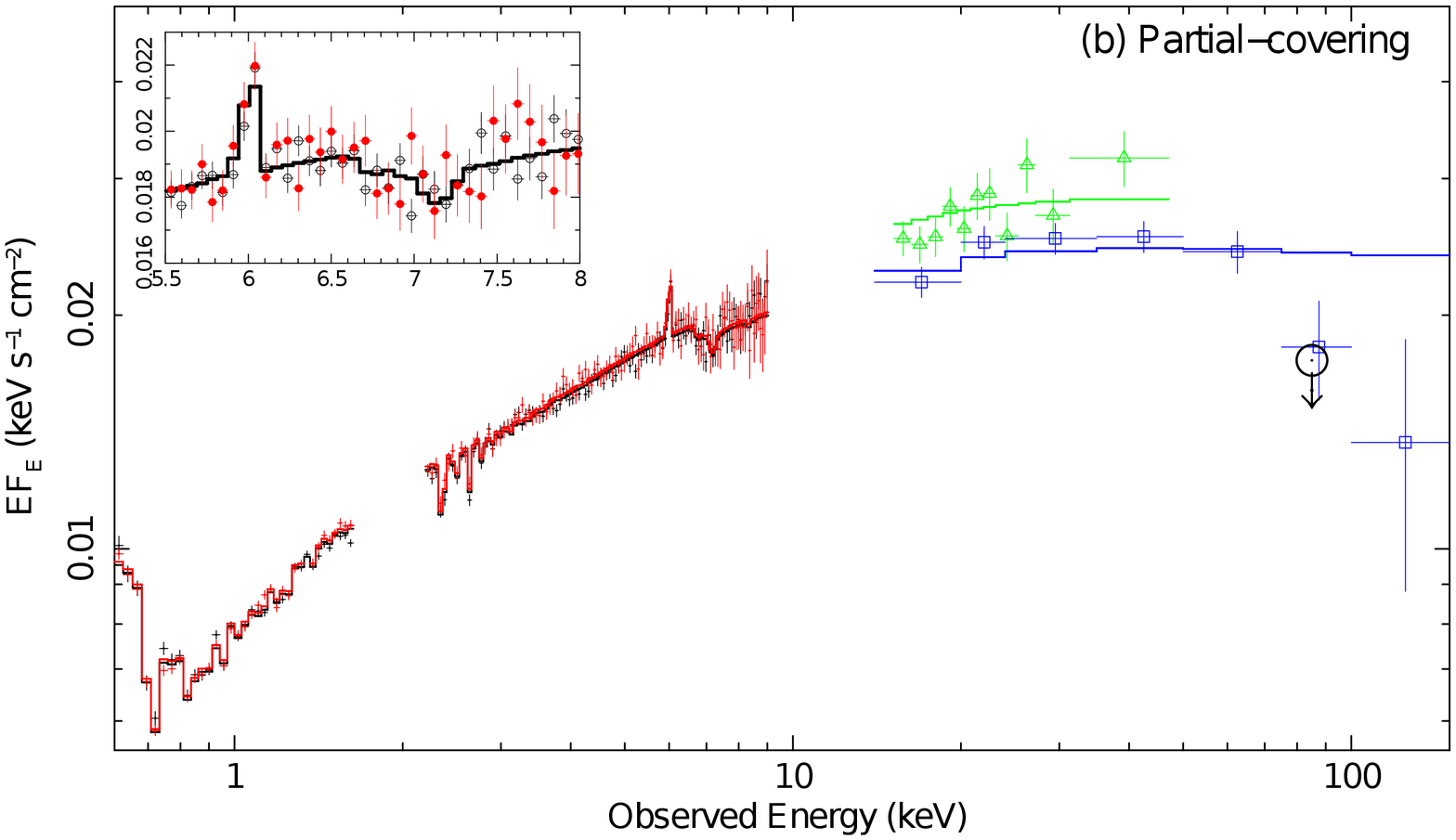}
\caption{Unfolded best-fit models for (a) the fully- and (b) the partially-covering models. Inset in each panel is an enlarged view of the Fe\,K region of each model. The $1\sigma$ upper limit to the HXD/GSO flux is represented by the open circle in both plots and is consistent with the data obtained from the {\it Swift}/BAT 58month  all-sky survey. Note that we have not included the tentative Fe\,{\sc xxvi} component from the inset panel of (a).}
\label{broadband_continuum}
\end{figure*}

While the overall fit statistic for the partial covering model is good ($\chi^{2}=335.6/308$; null prob=0.13) the fit still appears to overestimate the flux at the highest energies (see Figure \ref{broadband_continuum}, bottom panel). We thus tried to replicate the roll-over by adding an exponential roll-over using the {\tt highecut} model; however this only marginally improved the fit ($\Delta\chi^{2}=4.9$ for one additional free parameter). Nevertheless, the final fit statistic for the partial covering model is $\chi^{2}/\rm{dof}=330.7/307$ (null hypothesis=0.168) which is comparable to that obtained with the fully-covering model. 

\begin{table}[h]
\begin{center}
\caption{Best-fit parameters for the partial-covering model}
\begin{tabular}{c c c c}
\hline
\multicolumn{4}{c}{\textbf{Continuum parameters}} \\ \hline\hline
\multirow{2}{*}{Flux (erg\,cm$^{-2}$\,s$^{-1}$)} & $F_{\rm 0.5-2.0}$ & \multicolumn{2}{c}{$3.9\times10^{-11}$} \\
 & $F_{\rm 2.0-10.0}$ & \multicolumn{2}{c}{$5.3\times10^{-11}$} \\
\multirow{2}{*}{Luminosity (erg\,s$^{-1}$)} & $L_{\rm 0.5-2.0}$ & \multicolumn{2}{c}{$3.88\times10^{44}$} \\
 & $L_{\rm 2.0-10.0}$ & \multicolumn{2}{c}{$5.18\times10^{44}$} \\ 
\hline
\multicolumn{4}{c}{\textbf{Best-fit model components}} \\ [0.5ex]
Component & Parameter & Value & $\Delta\chi^{2\,h}$ \\
\hline\hline
Galactic absorption & $N_{\rm H}^{a}$ & $2.42\times10^{20}$ & --- \\ [1ex]
highecut & $E_{\rm cut}$ & $>244$ & 10.1 \\ [1ex]
power-law & $\Gamma$ & $1.96\pm0.01$ & -- \\ [1ex]
Reflection & Abund & 1$^{*}$ & 54.9$^{\,i}$ \\ [0.5ex]
 & $\xi^{e}$ & $<25$ & \\ [0.5ex]
 & R & $<0.2 $& \\ [1ex]
{\tt zxipcf} & $N_{\rm H,pcov}$ & $(1.2\pm0.2) \times10^{23}$ & 217.9 \\ [0.5ex]
 & $\log\xi_{\rm pcov}$ & $-0.11^{+0.21}_{-0.19}$ & \\ [0.5ex]
 & $f_{\rm cov}$ & $0.27^{+0.02}_{-0.03}$ & \\[0.5ex]
 & $v_{\rm turb}$ & 200  \\[1ex]
Soft X-ray absorber 1 & $N_{\rm H}^{a}$ & $(7.5\pm0.4)\times10^{21}$ & 110.3 \\ [0.5ex]
 (Zone 1) & $\log\xi^{e}$ & $1.79\pm0.02$ & \\ [0.5ex]
 & $v_{\rm out}^{f}$ & $<300$  \\ [0.5ex]
 & $v_{\rm turb}^{g}$ & $200$  \\ [1ex] 
Soft X-ray absorber 2 & $N_{\rm H}^{a}$ & $(2.1^{+0.1}_{-0.3})\times10^{21}$ & 106.3 \\ [0.5ex]
 (Zone 2) & $\log\xi^{e}$ & $1.83\pm0.04$ & \\ [0.5ex]
 & $v_{\rm out}^{f}$ & $27000\pm3000$  \\ [0.5ex] 
 & $v_{\rm turb}^{g}$ & 200\,km\,s$^{-1}$ \\ [1ex] 
Fe\,L absorber & $N_{\rm H}^{a}$ & $4.0^{+0.5}_{-0.6}\times10^{21}$ &  48.7 \\ [0.5ex]
(Zone 3) & $\log\xi^{e}$ & $3.01^{+0.23}_{-0.03}$ \\ [0.5ex]
 & $v_{\rm out}^{f}$ & $39000^{+9000}_{-6000}$  \\ [0.5ex]
 & $v_{\rm turb}^{g}$ & 5000  \\ [1ex]
S\,\textsc{xv} absorber & $N_{\rm H}^{a}$ & $6.4^{+4.3}_{-2.6}\times10^{21}$ & 51.6 \\ [0.5ex]
(Zone 4) & $\log\xi^{e}$ & $2.47^{+0.19}_{-0.24}$ & \\ [0.5ex]
 & $v_{\rm out}^{f}$ & $6300\pm5400$  \\ [0.5ex]
 & $v_{\rm turb}^{g}$ & 1000  \\ [1ex]
S\,\textsc{xvi} absorber & $N_{\rm H}^{a}$ & $>6.7\times10^{21}$ & 20.7 \\ [0.5ex]
(Zone 5) & $\log\xi^{e}$ & $3.37^{+0.18}_{-0.43}$ & \\ [0.5ex]
 & $v_{\rm out}^{f}$ & $<14000$  \\ [0.5ex] 
 & $v_{\rm turb}^{g}$ & 1000  \\ [1ex]
\hline
\label{final_pcov}
\end{tabular} \\
\end{center}
$^{a}$ Absorber column density in units of cm$^{-2}$. \\
$^{b}$ Cut-off energy in units of keV. \\
$^{e}$ Ionisation parameter in units of erg\,cm\,s$^{-1}$. \\
$^{f}$ Outflow velocity of the absorber given in units of km\,s$^{-1}$. \\
$^{g}$ Assumed absorber turbulent velocity in units of km\,s$^{-1}$. \\
$^{h}$ Significance of component with respect to the final best-fitting model.\\
$^{i}$ Statistical significance of {\sc reflionx} component is given in comparison to a fit where Fe\,K$\alpha$ is left unmodelled. \\
\end{table}

\section{discussion}
\label{discussion}

\subsection{The soft excess}
\label{soft-excess}
We have demonstrated that the soft excess in MR\,2251-178 can be adequately modelled either as an intrinsic component of the emergent X-ray continuum (i.e. an accretion disk black body), or as the side effect of an absorption dominated (partially covered) spectrum below 10\,keV. However, there are several further alternative interpretations which we also considered.

\subsubsection{Comptonisation}
A particular caveat of the fully-covering model is the manner in which we have parameterised the soft excess as being the thermal emissions from a geometrically thin accretion disk (\texttt{diskbb}), and thus as an intrinsic component of the X-ray spectrum. While simple and generally well fitting in a wide range of objects, such an interpretation typically requires a thermal continuum with a temperature higher than that expected for geometrically thin accretion disks (e.g. Shakura \& Sunyaev 1974; Porquet et al. 2004; Gierli\'nski \& Done 2004). The expected temperature for a geometrically thin disk is approximately $T(r)\approx6.3\times10^{5}(\dot M_{\rm acc}/\dot M_{\rm edd})^{1/4}M_{8}^{-1/4}(r/R_{s})^{-3/4}$\,K, where $M_{8}$ is the accretion  disk mass in units of $10^{8}\,M_{\odot}$. With ($\dot M_{\rm acc}/\dot M_{\rm edd})\sim0.15$ (see later) and $M_{8}\sim2.4$ (Dunn et al. 2008) appropriate for MR\,2251-178, and assuming most of the emission originates at the innermost stable circular orbit ($r_{\rm isco}=3R_{\rm s}$ for a Schwarzschild black-hole), we have $kT\approx20$\,eV. This is roughly 3 times cooler than the best-fitting \texttt{diskbb} temperature ($kT\sim62$\,eV) and implies that the soft excess cannot easily be attributed to thermal emission from a standard accretion disk without modification. 

An alternative interpretation of the soft excess as an intrinsic component is that it is the Wien tail of soft photons from the accretion disk which have been Comptonised by a hot corona located above the plane of the disk (e.g. Czerny \& Elvis 1987; Haardt \& Maraschi 1991, 1993; Nayakshin, Kazanas, \& Kallman 2000; Nayakshin 2001). If the distribution of electrons in such a Comptonising corona is ``quasi-Maxwellian'' (i.e., has both a thermal and non-thermal component; see Coppi 1999), it may be able to produce both the hard X-ray power-law continuum for which it is typically invoked, as well as the soft excess.

To see whether this interpretation was feasible, we replaced the \texttt{ diskbb} component in our best-fitting fully-covering \textsc{xstar} model with the \texttt{comptt} component (Titarchuk 1994). The thermal temperature of the input accretion disk photons was assumed to $kT\approx20\,$eV, while the redshift was fixed to that of the source. The resulting fit was comparable ($\Delta\chi^{2}/\rm{dof}=335.1/308$), and the other model parameters were all consistent with those obtained previously when using \texttt{diskbb}. The Comptonising electron temperature is $kT<5.3$\,keV and the optical depth of the corona is $\tau\sim0.40^{+0.14}_{-0.36}$.
\subsubsection{Ionised reflection}
Another alternative explanation for the soft excess in general is as a byproduct of ionised reflection from optically thick matter in the nuclear regions of the AGN, e.g. the accretion  disk, the broad line region (BLR) or the molecular the torus. However, due to the lack of a strong ionised reflection component and the absence of a broadened iron line, this seems unlikely in the case of MR\,2251-178.

\subsubsection{Section summary}
In summary, the soft excess in MR\,2251-178 is well modelled by either a thermal accretion  disk black body or as the result of the Comptonisation of soft thermal photons from the accretion disk. A partial covering interpretation of the dataset is also capable of replicating the soft-excess albeit with a steeper power-law photon index, i.e. $\Gamma\sim2.0$, which is similar to the value found commonly for radio-quiet quasars (e.g. Porquet et al. 2004). The lack of a strong ionised reflection component above 10\,keV and the absence of a broadened Fe\,K$\alpha$ line suggests than an origin through blurred reflection is unlikely. Regardless of the true nature of the soft excess in MR\,2251-178, the parameters that describe the fully-covering absorption components are largely invariant between the differing interpretations.

\subsection{The Fe\,K$\alpha$ line}
\label{feK-line}
From past observations of MR\,2251-178 there has been significant uncertainty in the reported Fe\,K$\alpha$ emission EW: i.e. $125^{+100}_{-105}$\,eV (\textit{Ginga}; Mineo \& Stewart 1993), $190^{+140}_{-95}$\,eV (\textit{ASCA}; Reynolds 1997), $62^{+12}_{-25}$\,eV, (\textit{BeppoSAX}; Orr et al. 2001) $53\pm20$\,eV (\textit{XMM-Newton}; Kaspi et al. 2004) and $23^{+14}_{-13}$\,eV (\textit{Chandra}; Gibson et al. 2005). 

A possible explanation for the uncertainty is that the previous observations lacked the spectral resolution and S/N in the Fe\,K band in order to resolve the absorption feature blue-ward of the Fe\,K$\alpha$, and this may have artificially broadened the measured EW of the Fe\,K emission line. Incidentally, we note that the two observations thus far which have reported the presence of Fe\,\textsc{xxv/xxvi} absorption (this work and Gibson et al. 2005) have also reported the tightest constraint on the Fe\,K$\alpha$ EW. 

Even so, the Fe\,K$\alpha$ EW we find for MR\,2251-178 is still below average. For example, based on an archival study of 36 sources in the {\it Chandra}/HETG archive (which includes MR\,2251-178) Shu et al. (2010) have reported the narrow core of the Fe K$\alpha$ emission line has a weighted mean EW of $70\pm4$\,eV, which is substantially broader than that which we find here. Furthermore, from their large {\it Suzaku} study of nearby Seyfert galaxies, Fukazawa et al. (2010) have found that a mean Fe\,K$\alpha$ EW of around $55$\,eV for sources which are absorbed by column densities similar to MR\,2251-178 (i.e. $N_{\rm H}\la10^{22}$\,cm$^{-2}$). Given that MR\,2251-178 is a highly X-ray luminous AGN, the apparent weakness of the Fe\,K$\alpha$ line could be a result to the `X-ray Baldwin Effect' - an apparent anti-correlation between source X-ray luminosity and the Fe\,K$\alpha$ EW, which has been well documented in the literature (e.g. Iwasawa \& Taniguchi 1993; Page et al. 2004; Jiang et al. 2006; see also Shu et al. 2010 and Fukazawa et al. 2010). Furthermore, given that MR\,2251-178 has a FR\,I radio morphology, it could also be due to a dilution of the observed X-ray spectrum by a component of X-ray continuum emitted in a radio-jet. The radio-loudness of an object is typically given in terms of the radio-loudness parameter, $R_{\rm L}$, defined by 
\begin{equation}
R_{\rm L}=\log\left(\frac{F_{5\,\rm{GHz}}}{F_{4400\rm{\AA}}}\right)
\end{equation}
where $F_{5\,\rm{Ghz}}$ is the monochromatic radio flux at 5\,Ghz and $F_{4400\rm{\AA}}$ is the optical flux at $\lambda=4400$\AA. In this regime, objects where $R_{\rm L}>1$ are regarded as radio-loud, while those where $R_{\rm L}<1$ are radio-quiet. For MR\,2251-178 Reeves \& Turner (2000) found that $R_{\rm L}=-0.43$ which suggests that any present radio-jet is very weak and unlikely to contribute in the X-ray band.

A further possibility is that the weakness of the Fe\,K$\alpha$ emission may be due to the inner accretion disk being shielded by the outflow. If this is the case, this could also explain the lack of a strong ionised reflection component as the observed reflection would have to originate at distances further from the central engine, i.e. the BLR or the torus. Sim et al. (2010) have also recently shown that Fe\,K$\alpha$ emission can originate via reflection from dense material in an outflowing wind (with EWs$\sim10$s of eV), which may also be applicable here. Furthermore, the Fe\,K$\alpha$ EW can also become suppressed in the presence of a sufficiently high-column absorber  (e.g. Miller et al. 2010; Yaqoob et al. 2010) such as those found here.

\begin{table*}
\begin{minipage}{130mm}
\caption{Summary of absorption zone parameters}
\begin{tabular}{@{}l c c c c c}
\hline
 & Soft X-ray 1 & Soft X-ray 2 & S\,\textsc{xv} & S\,\textsc{xvi} & Fe\,L/Fe\,K \\
 & (Zone 1) & (Zone 2) & (Zone 3) & (Zone 3) & (Zone 5) \\ \hline\hline
$v_{\rm out}$ (km\,s$^{-1}$) &$ 300^{*}$ & $300^{*}$ & $6300^{+5100}_{-4200}$ & $19800^{+4200}_{-5100}$ & $42000^{+3000}_{-5000}$ \\ [0.5ex]
$R$ (pc) & $2\times10^{6}$ & 700 & 300 & 30 & 100 \\ [0.5ex]
$R_{\rm esc}$ (pc) & 20 & 20 & 0.06 & 0.005 & $0.001$ \\ [0.5ex]
$\dot M_{\rm out}$ ($\rm M_{\odot}$\,yr$^{-1}$) & $33000b$ & $100b$ & $1500b$ & $600b$ & $2700b$ \\ [0.5ex]
$b$ & $10^{-5}$ & $10^{-3}$ & $10^{-2}$ & $10^{-3}$ & $10^{-3}$\\ [0.5ex]
$\dot E_{\rm out}$ (erg\,s$^{-1}$) & $\sim10^{45}b$ & $\sim10^{42}b$ & $\sim10^{46}b$ & $\sim10^{46}b$ & $\sim10^{48}b$ \\ [0.5ex]
$\dot E_{\rm out(corr)}$ (erg\,s$^{-1}$) & $\sim10^{40}$ & $\sim10^{39}$ & $\sim 10^{44}$ & $\sim 10^{43}$ & $\sim10^{45}$ \\ [1ex]
\hline
\label{absorbers}
\end{tabular}
$^{*}$Denotes the parameter was fixed at this value. \\
\end{minipage}
\end{table*}

\subsection{The multi-component outflow of MR\,2251-178}
\label{quantifying-the-outflow}
We have shown that the X-ray spectrum of MR\,2251-178 can be adequately modelled with both a fully-covering and an absorption dominated partially-covered interpretation. We find that both of these models require the presence of a number of fully-covering absorbers to account for the observed Fe UTA, Fe\,L (Fe\,{\sc xxiii-xxiv}), S\,{\sc xv}, S\,{\sc xvi} and Fe\,{\sc xxv-xxvi} absorption lines. In this section, we turn our attention to these absorbers and investigate their geometries, kinematics, and, ultimately, estimate their energetic output. For simplicity we consider here the absorbers as per the fully-covering model, and focus on the Fe\,L absorber as it is likely to be the most energetic given its high $v_{\rm out}$. For completeness, however, the methods discussed here were individually applied to all absorbers in the fully-covering model and the results are summarised in Table \ref{absorbers}.

Note that as the parameters of the Fe\,L absorber are largely invariant between the two continuum models, the results obtained here are applicable to either interpretation and are thus not model dependent. 

\subsubsection{Distance}
\label{absorber-distance}
The observed column density of a spherically symmetric absorber a distance $R$ from the ionising source is equal to the absorber density multiplied by the thickness of the shell of material; i.e. $N_{\rm H}\approx n\delta R$. From the definition of the ionisation parameter, $\xi=L_{\rm ion}/nR^{2}$,
where $L_{\rm ion}$ is the ionising luminosity between 1-1000 Rydbergs (13.6\,eV to 13.6\,keV) and $n$ is the hydrogen gas density, this then implies that the inner face of the absorber, $R$, is located at
\begin{equation}
R = \frac{L_{\rm ion}}{\xi\,N_{\rm H}}\left(\frac{\delta R}{R}\right)
\label{explicitdist}
\end{equation}
from the central ionising continuum. Assuming that the absorber forms a thin shell where $\delta R/R < 1$, equation (\ref{explicitdist}) then sets an upper limit on the distance to the inner face of the absorbing material, i.e. $R < L_{\rm ion}/\xi\,N_{\rm H}$. From the best-fit continuum parameters of the fully-covering model, $L_{\rm ion}$ is estimated to be of the order of $2\times10^{45}$\,erg\,s$^{-1}$. Therefore, taking $N_{\rm H}\sim4.4\times10^{21}$\,cm$^{-2}$ and $\log\xi/\rm{erg\,cm\,s}^{-1}\sim3.3$ characteristic of the Fe\,L absorber, we find $R\la100$\,pc.

Assuming that the outflow escapes the system we can set a lower limit on the radial distance by considering the escape radius of the material. For a spherical geometry the radius at which material will be able to escape the BH is given by
\begin{equation}
R_{\rm esc}\geq\frac{2GM}{v_{\rm out}^{2}}\approx\left(\frac{2c^{2}}{v_{\rm out}^{2}}\right)R_{\rm g}
\end{equation}
with $R_{\rm g}\equiv GM/c^{2}$ being the gravitational radius. Given that the Fe\,L absorber has an outflow velocity of $\sim0.14$\,c, $R_{\rm esc}\gtrsim100\,R_{\rm g}$. This suggests that the outflow may have an origin consistent with an accretion disk wind or the BLR. Indeed, Kaspi et al. (2004) set a lower limit on the size of the BLR in MR\,2251-178 of $0.04\pm0.01$\,pc ($\approx 3000$\,Rg), which has significant overlap with the upper limit obtained here.

\subsubsection{Mass outflow rates}
\label{outflow-rates}
We can also estimate the mass outflow rate. Assuming a quasi-spherical system:
\begin{eqnarray}
\dot M_{\rm out}&=&4\pi b R^{2} n_{e} m_{p} v_{\rm out} \\
	&\approx&4\pi b \left(\frac{L_{\rm ion}}{\xi}\right) m_{p} v_{\rm out}
\end{eqnarray}
where $b\equiv\Omega/4\pi\leq1$ is the geometrical factor allowing for an outflowing wind subtending sold angle $\Omega$. The parameters of the Fe\,L absorber suggest that it is capable of expelling $\sim2700b\,M_{\odot}$\,yr$^{-1}$. For comparison, the accretion rate required to maintain the observed radiative bolometric output is:
\begin{equation}
\dot M_{\rm acc}=\frac{L_{\rm bol}}{\eta c^{2}}
\end{equation}
where $L_{\rm bol}$ is the bolometric luminosity and $\eta$ is the accretion efficiency. With $\eta=0.06$ for a Schwarzschild black-hole accreting at maximum efficiency, and adopting a UV-determined bolometric luminosity of $4.3\times10^{45}$\,erg\,s$^{-1}$ (Dunn et al. 2008), we find that the AGN in MR\,2251-178 requires an accretion rate of $\dot M_{\rm acc}\sim1.3\,M_{\odot}$\,yr$^{-1}$ to maintain its bolometric radiative output. This is orders of magnitude lower than the estimated mass outflow rate and implies that either, (1) the AGN will exhaust its fuel supply of accreting material over a relatively short timescale in order to maintain an effective mass outflow rate much higher than $\dot M_{\rm acc}$, or (2) the outflow is not spherical and has a small covering fraction or is rather clumpy (i.e. $b\ll1$). Similar conclusions were also reached by Gibson et al. (2005) in their analysis of the {\it Chandra}/HETG observation.

By assuming that $\dot M_{\rm out} \la \dot M_{\rm acc}$ we can estimate the clumping/covering $b$ factor of the outflow with respect to the central nucleus, i.e. $b\approx\dot M_{\rm acc}/\dot M_{\rm out}\approx10^{-3}$. The resulting inferred covering factors are listed in Table \ref{absorbers}. We note that in principle $b$ could be larger than this. However, this estimate allows us to set a conservative lower-limit to the energetics of the outflow in the next section.

\subsubsection{Energetics}
\label{outflow-energetics}
The kinetic luminosity of the Fe\,L absorber is:
\begin{equation}
\dot E_{\rm out}=\frac{1}{2}\dot M_{\rm out}v_{\rm out}^{2}b\sim10^{48}b\,\rm{erg\,s}^{-1}
\end{equation}
which implies a conservative mechanical output of $\sim10^{45}$\,erg\,s$^{-1}$ for $b\sim10^{-3}$ as discussed above. This makes the Fe\,L absorber the most significant outflowing component in terms of overall energetics, despite its substantial putative clumping. The energetics of the other absorbers are listed in Table \ref{absorbers}. 

Assuming that the BH mass grows through accretion alone, we can then estimate a required `active phase' for the QSO as the current estimated BH  mass ($2.4\times10^{8}\,M_{\odot}$) divided by the accretion rate, i.e. $\sim10^{8}$\,yr. Combining this with the mechanical output calculated previously suggests that the Fe\,L absorber has an estimated mechanical output of $\sim10^{60}$\,erg over the lifetime of the AGN. This is comparable to the binding energy of a typical $10^{11}\,M_{\odot}$ galaxy bulge and further supports the premise that such outflows may be energetically significant in AGN/Galaxy feedback scenarios. 

\section{Summary}
\label{summary}
We have performed a broad-band spectral analysis of the radio-quiet quasar MR\,2251-178 using recent observations from \textit{Suzaku} and the \textit{Swift} BAT. Below is a summary of our results:
\begin{itemize}
\item The continuum shows considerable curvature above around $\sim 10$\,keV and an apparent soft excess is present below 1\,keV. A weak ($EW=26^{+16}_{-6}$\,eV), unresolved ($\sigma < 117$\,eV), Fe\,K$\alpha$ emission line from neutral material is present at $\sim6.4$\,keV. The spectral curvature and soft excess can be equally well modelled as an artefact of an absorption dominated, partially covered, continuum or as a fully covered intrinsic continuum component.

\item Absorption lines due to the Fe\,UTA and the Fe\,L-shell are present below 2\,keV. The Fe\,L line is consistent with being a blueshifted blend of 2s$\rightarrow$3p transitions from Fe\,\textsc{xxiii} and Fe\,\textsc{xxiv}. Two further absorption lines are detected at $\sim2.5$\,keV and $\sim2.8$\,keV, which we conservatively identify with S\,\textsc{xv} He$\alpha$ and S\,\textsc{xvi} Ly$\alpha$. For this identification the lines require blueshifted velocities of $\sim0.02$\,c and $\sim0.07$\,c, respectively. From Montecarlo simulations we find that all soft X-ray absorption lines are significant at the $>99.9$\% level.

\item We also detect significant ($99.3\%$ from MC) absorption in the Fe\,K band ($E\sim7.5$\,keV; rest-frame) which is consistent with the presence of blueshifted Fe\,\textsc{xxv-xxvi}. A single high turbulence ($v_{turb}=5000$\,km\,s$^{-1}$) ionised absorber provides a good fit to both the Fe\,L-shell line at $\sim1.29$\,keV and to a portion of the absorption present in the Fe\,K band, requiring an outflow velocity of $v_{\rm out}\sim0.14$\,c.

\item We derive distances to the absorbers and find that the Fe\,L absorption may originate at sub-parsec scales, possibly in an accretion disk wind. By balancing the mass outflow and mass accretion rates, we estimate covering fractions (or `clumpiness') and subsequent energetics for all absorbers. The Fe\,L absorber is consistent with being substantially clumped, by a factor of around $10^{-3}$, but still dominates in terms of the overall energetics with an estimated kinetic output of $\sim10^{45}$\,erg\,s$^{-1}$.
\end{itemize}

\section*{Acknowledgments}
The {\it Suzaku} X-ray observatory is a collaborative mission between the Japan Aerospace Exploration Agency (JAXA) and the National Aeronautics and Space Administration (NASA). This research has made use of the NASA/IPAC Extragalactic Database (NED) which is operated by the Jet Propulsion Laboratory, Caltech, under contract with the National Aeronautics and Space Administration. We would like to thank the anonymous referee for their comments on this work, and Dean McLaughlin for his numerous helpful suggestions. JG and VB would like to acknowledge support from the UK STFC research council.

\appendix

\section{Line identifications}
\label{line-identifications}
The outflow velocities (and resultant energetics) of the absorbers are contingent on the correct identifications of the absorption features. Most of the absorption features detected in the \textit{Suzaku} spectrum have rest-frame energies which do not correspond to any strong features expected in the X-ray band, and have thus been conservatively identified as being the blueshifted absorption from the strongest expected nearby spectral transition. Here we discuss alternative identifications for these absorption lines.

\subsection{Fe\,{\sc xxiv}}
The observed energy range of the suspected Fe\,\textsc{xxiv} (2s$\rightarrow$3p) line, i.e. 1.28-1.30\,keV, is populated by an array of high-order transitions from low-abundance elements which makes a rest-frame identification unlikely. For example, the 2s$\rightarrow$$n$d transitions of Co\,\textsc{xviii} and Co\,\textsc{xx} are found in this energy range which can be ruled out based on very low astrophysical abundances (i.e. Co/H $\sim 10^{-8}$). Indeed, these transitions are also unlikely as they are expected to be very weak in comparison to their lower-order counterparts.

The line at $1.29$\,keV is also unlikely to be identified with any blueshifted lines of the expected abundant elements (i.e. O, Ne) without an extreme outflow velocity. Indeed, if identified with H- or He-like Neon (1.02\,keV and 0.922\,keV, respectively) the outflow velocity would be $\sim0.3-0.4$\,c. This is much larger than that typically found in AGN warm absorbers (e.g. Blustin et al. 2005). Furthermore, an identification with the 1s$\rightarrow$2p lines of Na\,{\sc x/xi} (1.12\,keV and 1.24\,keV) is unlikely as such Na absorption has not been reported in even the longest exposure AGN X-ray spectra (e.g. Kaspi et al. 2002).

As an identification with any local 1s$\rightarrow$2p transition is unlikely, we also considered nearby 2s$\rightarrow$3p transitions. The most likely 2s$\rightarrow$3p transition local to the observed $\sim$1.29\,keV line is that of Fe\,\textsc{xxiv} at $\sim1.165$\,keV. For this identification the line would then require an outflow velocity of $\sim0.1$\,c. We note that while the oscillator strengths of the 2s$\rightarrow$3p lines of Fe XXV (expected at $E~1.22-1.24$\,keV) are expected to be higher than those of Fe XXIV in this band, the required outflow velocity assuming this identification ($v_{\rm out}~0.04-0.06$\,c) would be incompatible with that obtained from {\sc xstar} fitting (see \S\ref{fel_abs}).

\subsection{ S\,{\sc xv/xvi}}
The lines at $E=2.52\pm0.02$\,keV and $E=2.79\pm0.03$\,keV do not correspond with any expected transitions in the rest-frame. The only transitions which have consistent rest-frame energies are those associated with extremely low abundance elements which are not expected to be observed in the X-ray spectrum (e.g. Kr, Rb, Mo).

The higher energy line is consistent with the 1s$\rightarrow$3p He-like Sulphur line at $E\sim2.88$\,keV but such an identification is again unlikely given that at the best-fitting ionisation stage (Table \ref{final_fcov}) H-like Sulphur is expected to be the dominant ion (Kallman \& McCray 1982). Furthermore, if the line was caused by the He-like Sulphur 1s$\rightarrow$3p line, we would still expect to see much stronger 1s$\rightarrow$2p transitions, which are not present. Thus a $v_{\rm out}=0$ origin appears to be ruled out for both of the lines and they are most likely be blueshifted.

The most likely ions in the 2-3\,keV energy band which are expected to be strong in the X-rays are He- and H-like Sulphur at $E=2.461$\,keV and $E=2.623$\,keV, respectively. An identification with Silicon (e.g. Si\,\textsc{xiv} Ly$\alpha$ at 2.01\,keV) would require a substantially larger soft X-ray outflow velocity $v\sim0.2-0.4$\,c and thus seems unlikely; as does an identification with any lower Z ions for similar reasons.

\subsection{Fe\,{\sc xxv/xxvi}}
While their oscillator strengths suggest that the absorption features between 7-8\,keV are most likely Fe\,\textsc{xxv/xxvi}, several other elements are also expected to have transitions above $\sim7$\,keV which may interfere with this identification. Most notable of these are H- and He-like Nickel which have 1s$\rightarrow$2p lines at rest-frame energies of $\sim7.8$\,keV and $\sim8.1$\,keV, respectively. The observed energy of the absorption feature is consistent with neither of these transitions without significant redshift, however, making an identification unlikely. An identification with lower ionisation Nickel could be plausible but the column densities at which moderately ionised Nickel would be observed are unacceptably high (i.e. $N_{\rm H}>10^{24}$\,cm$^{-2}$) and with an abundance ratio of Fe/Ni$\sim20$ (Grevesse \& Suvaul 1998) we would also expect to observe hugely dominant 1s$\rightarrow$2p Fe transitions, which are not observed. Thus an identification with Ni would appear to be ruled out.

For the tentative high-ionisation Fe\,K absorber an identification with any ions other than Fe\,\textsc{xxvi} is unlikely given the high ionisation as Fe\,\textsc{xxvi} is expected to be the only abundant ions at the best fit ionisation parameter (Kallman et al. 2004). Furthermore, while Zone 5 may be contributing Fe\,{\sc xxiii-xxiv} 1s$\rightarrow$2p lines to the Fe\,K band, Fe\,\textsc{xxv} is expected to be the dominant ion at the measured ionisation parameter and we adopt it as our most probable line identification for the contribution of Zone 5 to the Fe\,K band.

\label{lastpage}

\end{document}